\documentclass[pdftex, aps, prd, letterpaper, superscriptaddress, nofootinbib, twocolumn,
               preprintnumbers, longbibliography]{revtex4-1}
\pdfoutput=1
\usepackage{amsmath,amssymb,amsbsy}
\usepackage[utf8]{inputenc}
\usepackage{graphicx,stix}
\usepackage{soul,xcolor,topcapt}
\usepackage[colorlinks=true, linkcolor=blue, filecolor=blue, urlcolor=blue, citecolor=blue, pdftex, plainpages=false]{hyperref}

\newcommand{\gGF}{g^2_{\textrm{GF}}}
\newcommand{\betaGF}{\beta_{\textrm{GF}}}
\newcommand{\LambdaGF}{\Lambda_{\textrm{GF}}}

\newcommand{\vev}[1]{\langle #1\rangle}
\newcommand{\svev}[1]{\left\langle #1\right\rangle}
\newcommand{\MSbar}{\ensuremath{\overline{\textrm{MS}} } }

\begin{document}
\preprint{ FERMILAB-PUB-23-096-V ~~SI-HEP-2023-04}
\title{\texorpdfstring{$\Lambda$}{Lambda} parameter of the SU(3) Yang-Mills theory from the continuous \texorpdfstring{$\beta$}{beta} function}

\author{Anna Hasenfratz}
\email{anna.hasenfratz@colorado.edu}
\affiliation{Department of Physics, University of Colorado, Boulder, Colorado 80309, USA}
\author{Curtis T.~Peterson}
\email{curtis.peterson@colorado.edu}
\affiliation{Department of Physics, University of Colorado, Boulder, Colorado 80309, USA}
\author{Jake van Sickle}
\affiliation{Department of Physics, University of Colorado, Boulder, Colorado 80309, USA}
\author{Oliver Witzel}
\email{oliver.witzel@uni-siegen.de}
\affiliation{Center for Particle Physics Siegen, Theoretische Physik 1, Naturwissenschaftlich-Technische Fakult\"at, Universit\"at Siegen, 57068 Siegen, Germany}

\begin{abstract}
Nonperturbative determinations of the renormalization group $\beta$ function  are essential to connect lattice results to perturbative  predictions of strongly coupled gauge theories and to  determine the $\Lambda$ parameter or the strong coupling constant. The continuous $\beta$ function  is very well suited for this task  because it is applicable both in the weakly coupled deconfined regime as well as the strongly coupled confined regime. Here we report on our results for the $\beta$ function of the pure gauge SU(3) Yang-Mills theory in the gradient flow  scheme. Our calculations cover the renormalized coupling range $\gGF \sim $1.2$- 27$, allowing for a direct determination of $\sqrt{8t_0} \Lambda_{\MSbar}$ in this system. 
Our prediction, $\sqrt{8 t_0} \Lambda_{\MSbar}=0.622(10)$, is in good agreement with recent direct determinations of this quantity.
\end{abstract}
\maketitle

\section{Introduction} 

The  renormalization group (RG) $\beta$ function is defined as the logarithmic derivative of the renormalized running coupling $g^2$ with respect to an energy scale $\mu$
\begin{equation}\label{eqn:def_beta_function}
\beta\big(g^2\big) = \mu^2\frac{\mathrm{d}g^2(\mu)}{\mathrm{d}\mu^2},
\end{equation}
and describes the  scale dependence  of the renormalized coupling of a 4-dimensional gauge-fermion system. 
 Precise determination of the $\beta$ function is essential to understand the nonperturbative running of the gauge coupling and to predict, e.g., the $\Lambda$ parameter in quantum chromodynamics (QCD)
\begin{multline}\label{eqn:lambda_parameter}
    \frac{\Lambda^2}{\mu^2}=\big(b_0 g^2(\mu)\big)^{-\frac{b_1}{b_0^2}}\exp\bigg(-\frac{1}{b_0 g(\mu)^2}\bigg) \\
    \times  \exp\bigg[-\int_{0}^{g^2(\mu)}\!\!\!\!\mathrm{d}x \;\bigg(\frac{1}{\beta(x)}+\frac{1}{b_0 x^2}-\frac{b_1}{b_0^2 x}\bigg)\bigg],
\end{multline}
where $g^2(\mu)$ is  the renormalized coupling at  the
energy scale $\mu$ and the constants $b_0$ and $b_1$ denote the universal 1- and 2-loop coefficients of the RG $\beta$ function. Beyond 2-loop the $\beta$ function is scheme dependent, but it is straightforward to connect the $\Lambda$ parameter obtained using a different scheme to the phenomenologically favored $\overline {\mathrm{MS}}$ scheme 
using the 1-loop perturbative relation between the corresponding running couplings. 

In the chiral limit of QCD, the $\Lambda$ parameter sets the physical scale. Its accurate determination allows the prediction of the strong coupling constant $\alpha_{s}(\mu)$ at $\mu \geq M_Z$, the mass of the $Z$ boson \cite{FlavourLatticeAveragingGroupFLAG:2021npn, Bruno:2017gxd}. $\Lambda$ parameters of systems with different flavor numbers can be connected by nonperturbative decoupling relations. In particular, it is possible to predict $\alpha_{s}(M_Z)$ in $N_f=3$ QCD from the pure Yang-Mills $\Lambda$ parameter \cite{DallaBrida:2022eua}. A precise calculation of the $\Lambda$ parameter in the pure Yang-Mills system is therefore desirable. 

The RG transformation of an asymptotically free system has an ultraviolet (UV) fixed point (FP) on the $g^2_0=0$ critical surface, where $g_{0}^2$ denotes  the (marginally) relevant bare coupling. The UVFP and the renormalized trajectory (RT) emerging from the UVFP  describe continuum physics. The renormalized gauge coupling in Eq.~(\ref{eqn:def_beta_function}) ``measures'' the flow along the RT.
The gradient flow  (GF) transformation  \cite{Narayanan:2006rf, Luscher:2009eq, Luscher:2010iy} is a particularly promising choice to define a renormalization scheme, determine the running coupling $\gGF(\mu)$, and calculate a nonperturbative $\beta$ function using lattice simulations.
While on its own the GF transformation does not describe an RG transformation  because it does not reduce the degrees of freedom of the system, it can be modified to  define a complete RG scheme if a coarse-graining step is incorporated when defining expectation values \cite{Carosso:2018bmz, Carosso:2019qpb}. In this setup, the energy scale is set asymptotically by the GF flow time $t$
\begin{align}
  \mu \propto 1/\sqrt{8 t}.
\end{align}
For a generic RG flow, there is a one-to-one correspondence between running renormalized coupling and local observables that do not  renormalize \cite{Makino:2018rys}. In the GF scheme the renormalized  coupling
can be defined in terms  of the continuum flowed  energy density $E(t)$ as
\begin{equation}\label{eqn:inf_vol_gf_cpling}
\gGF(t) \equiv \mathcal{N} \svev{t^2 E(t)},
\end{equation}
since $E(t)$ has no anomalous dimension \cite{Luscher:2010iy}.
The normalization factor $\mathcal{N}={128\pi^2}/(3N_c^2-3)$ in Eq.~(\ref{eqn:inf_vol_gf_cpling}) is chosen  such that the renormalized coupling $\gGF$ 
matches the renormalized coupling of the $\overline {\mathrm{MS}}$ scheme at tree-level.

The GF coupling offers a convenient  low-energy hadronic scale  $1/\sqrt{8t_0}$,  with the flow time $t_0$  defined where the GF coupling takes a specified value, $\gGF(t_0) \equiv \mathcal{N} \times 0.3$ \cite{Luscher:2010iy}.  The $t_0$ scale has been used extensively in simulations to set the lattice scale and its continuum value is known in Yang-Mills systems \cite{Luscher:2010iy, DallaBrida:2019wur} and in QCD \cite{FlavourLatticeAveragingGroupFLAG:2021npn}
with 2+1 flavors \cite{Borsanyi:2012zs,RBC:2014ntl,Bruno:2016plf} and 2+1+1 flavors \cite{Dowdall:2013rya,MILC:2015tqx,Miller:2020evg}.  Further, we can use $\gGF$ to define  the RG $\beta$ function in the GF scheme  $\betaGF(\gGF)$.  If $\betaGF$ is determined up to $\gGF(t_0)\approx 15.8$,
we can obtain $\sqrt{8t_0} \LambdaGF$ by integrating Eq.~(\ref{eqn:lambda_parameter}) up to $\gGF(t_0)$.

The finite volume step-scaling function \cite{Luscher:1993gh,Fodor:2012td,DallaBrida:2019wur} is a commonly applied method to determine the $\beta$ function.  However, it requires that the lattice size is the only dimensionful quantity of the system, preventing its application in the confining, chirally broken regime of QCD. Since the $t_0$ scale corresponds to a low energy hadronic scale in the confining regime,  the step-scaling method cannot determine the $\beta$ function at strong enough gauge couplings to directly predict the $\Lambda$ parameter at the $t_0$ scale. An alternative approach is to determine the infinite volume continuous $\beta$ function as described in Refs.~\cite{Hasenfratz:2019hpg,Hasenfratz:2019puu,Fodor:2017nlp}. While this method requires separate infinite volume and continuum limit extrapolations, it is applicable in the confining regime.

In this paper, we report on our  findings on the continuous $\beta$ function up to and even beyond the renormalized gauge coupling $\gGF(t_0)$ in the pure gauge SU(3) Yang-Mills theory. Using the continuous $\beta$ function method we are able to determine the scale dependence of the running coupling in the confining regime. We do not need  any other hadronic observable to determine the $\Lambda$ parameter at the $t_0$ scale.  
We  compare our value of the $\Lambda$ parameter  to other determinations from Refs.~\cite{DallaBrida:2019wur,Wong:2023jvr} and values reported in the FLAG 2021 review \cite{FlavourLatticeAveragingGroupFLAG:2021npn}. 
Preliminary results of our work were reported in Ref.~\cite{Peterson:2021lvb}.  A similar calculation of the continuous $\beta$ function of SU(3) Yang-Mills theory  reports preliminary results in Ref.~\cite{Wong:2023jvr}.

 This paper is organized as follows. 
In Sec.~\ref{sec:num_dets} we summarize the details of our numerical simulations and  explain our determination of the continuous $\beta$ function in Sec.~\ref{sec:lat_g2}.  In the subsequent section we discuss the determination of the $\Lambda$ parameter and close by discussing our results in Sec.~\ref{sec:discussion}.

\section{Numerical Details}\label{sec:num_dets} 
\begin{table*}[tb]
\centering
\begin{tabular}{c@{\extracolsep{4pt}~~~~}cc@{~~~}cc@{~~~}cc@{~~~}cc@{~~~}cc}
  \hline\hline
   &\multicolumn{10}{c}{$L/a$} \\
  \cline{2-11}
  & \multicolumn{2}{c}{20}     & \multicolumn{2}{c}{24}      & \multicolumn{2}{c}{28}      & \multicolumn{2}{c}{32}      &  \multicolumn{2}{c}{48} \\
 \cline{2-3}  \cline{4-5}  \cline{6-7}  \cline{8-9}  \cline{10-11} 
  $\beta_b$ & Acceptance & No.	 & Acceptance & No.	 & Acceptance & No.	 & Acceptance & No.	 & Acceptance & No. \\ \hline
4.30 & 87.9\% & 451     & 86.6\% & 467  & 85.3\% & 297  & 80.7\% & 165 & $\cdots$ & $\cdots$ \\
4.35 & 86.5\% & 451     & 84.8\% & 458  & 82.0\% & 277  & 78.8\% & 171 & $\cdots$ & $\cdots$ \\
4.40 & 86.6\% & 451     & 80.8\% & 460  & 83.7\% & 272  & 82.3\% & 167 & $\cdots$ & $\cdots$ \\
4.50 & 85.1\% & 451     & 84.2\% & 501  & 86.2\% & 1391 & 81.0\% & 250 & $\cdots$ & $\cdots$ \\
4.60 & 86.1\% & 451     & 85.2\% & 490  & 83.3\% & 1040 & 84.9\% & 202 & $\cdots$ & $\cdots$ \\
4.70 & 84.2\% & 451     & 84.1\% & 490  & 80.5\% & 681  & 82.1\% & 201 & $\cdots$ & $\cdots$ \\
4.80 & 86.5\% & 451     & 88.0\% & 469  & 80.5\% & 681  & 78.9\% & 140 & $\cdots$ & $\cdots$ \\
4.90 & 85.0\% & 451     & 85.3\% & 491  & 82.7\% & 701  & 83.4\% & 163 & $\cdots$ & $\cdots$ \\
5.00 & 82.6\% & 451     & 85.5\% & 456  & 81.0\% & 772  & 77.3\% & 211 & 80.8\% & 124 \\ \hline
{\it 5.30} & {\it 84.4\%} & {\it 451}     & {\it 88.3\%} & {\it 534}  & {\it 82.9\%} & {\it 911}  & {\it 78.4\%} & {\it 656} &  {\it 81.8\%} & {\it 139} \\
5.50 & 83.6\% & 451     & 87.6\% & 456  & 81.8\% & 701  & 77.8\% & 608 & 78.2\% & 149 \\
6.00 & 84.4\% & 451     & 84.6\% & 476  & 84.6\% & 661  & 79.2\% & 472 & 76.8\% & 227 \\
6.50 & 81.1\% & 451     & 80.7\% & 486  & 82.8\% & 661  & 85.0\% & 563 & 77.4\% & 233 \\
7.00 & 81.3\% & 451     & 79.2\% & 461  & 81.7\% & 701  & 84.6\% & 527 & 74.7\% & 241 \\
7.50 & 82.6\% & 451     & 81.3\% & 466  & 80.5\% & 661  & 83.7\% & 489 & 73.6\% & 224 \\
8.00 & 81.3\% & 451     & 78.3\% & 456  & 76.1\% & 701  & 85.0\% & 487 & 73.3\% & 211 \\
8.50 & 78.8\% & 451     & 77.4\% & 461  & 79.5\% & 661  & 81.6\% & 462 & 74.6\% & 211 \\
9.00 & 78.2\% & 451     & 76.8\% & 581  & 78.0\% & 524  & 81.6\% & 531 & 71.7\% & 208 \\
9.50 & 77.4\% & 621     & 77.5\% & 481  & 77.7\% & 547  & 81.7\% & 541 & 69.2\% & 208 \\
\hline\hline
\end{tabular}
\caption{ The number of gauge field configurations analyzed and HMC acceptance rates for each bare coupling $\beta_b$ and volume $(L/a)^4$. Configurations are separated by 20 MDTUs. All ensembles with $\beta_b > 5.5$ are in the deconfined phase, while most with  $\beta_b\le 5.50$ either transition between confined and deconfined or are confined. The values for $\beta_b=5.30$ are set in italics because these ensembles are not part of the main analysis and are only used to estimate systematic effects. } 
 \label{table:ensembles}
 \end{table*}

\begin{figure}[tb]
    \centering
    \includegraphics[width=\columnwidth]{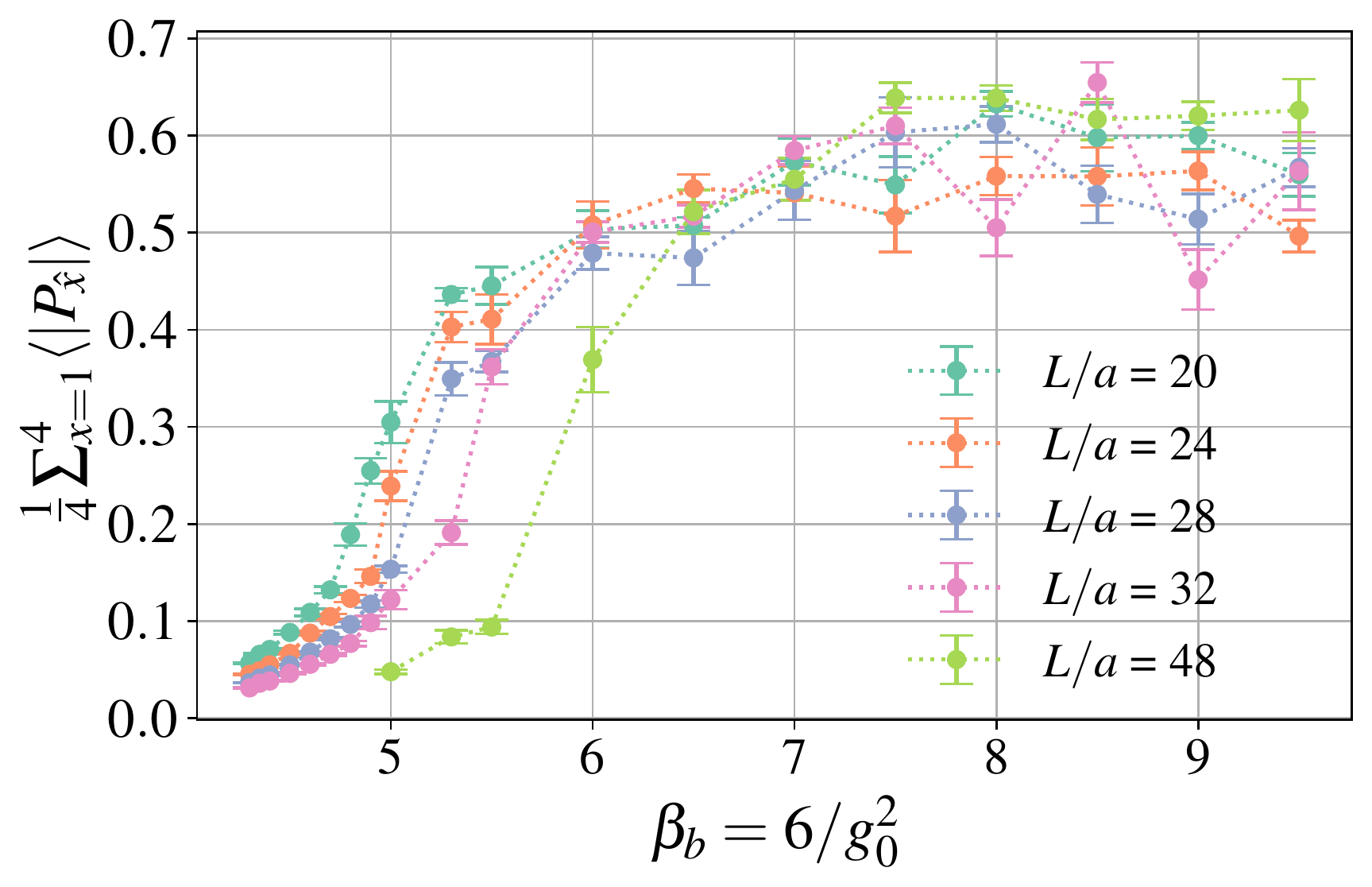}
    \caption{The expectation value of the Polyakov loop on our ensembles. We measure the Polyakov loop at flow time $t/a^2 = (L/a)^2/32$, and average the absolute values over all four directions.} 
    \label{fig:poly_loops}
\end{figure}

Our study is based on simulations performed using the tree-level improved Symanzik (L\"uscher-Weisz) gauge action \cite{Luscher:1984xn, Luscher:1985zq}. We consider nineteen bare gauge couplings 
and generate configurations with periodic boundary conditions (BC) in all four directions using the hybrid Monte Carlo (HMC) update algorithm \cite{Duane:1987de} as implemented in \texttt{GRID} \cite{Boyle:2015tjk,GRID}. We set the trajectory length to $\tau=2$ molecular dynamics units (MDTU) and save configurations every 10 trajectories (20 MDTU).  Subsequently, we use \texttt{QLUA} \cite{Pochinsky:2008zz,qlua} to perform gradient flow measurements.

Table \ref{table:ensembles} lists the number of thermalized configurations analyzed for each bare coupling and volume as well as the HMC acceptance rates which all range between 70\% and 90\%. 
In the strong coupling regime\footnote{To better distinguish between the RG $\beta$ function and the bare gauge coupling, we refer to the latter as $\beta_b\equiv 6/g_0^2$.} ($\beta_b \le 4.9$)  we have four volumes,  ($(L/a)^4=20^4, 24^4, 28^4, 32^4$), while at weaker couplings  ($\beta_b \ge 5.0$) we add a fifth, $48^4$ volume.\footnote{Although $L/a=20$ ensembles were generated for all $\beta_b$, our analysis uses $L/a=20$ only for $\beta_b\le 5.00$.}

The Polyakov loop expectation value in Fig.~\ref{fig:poly_loops} shows that most ensembles for $\beta_b\ge 5.50$ are in the deconfined phase, while most  $\beta_b \le 5.0$ ensembles are confining.\footnote{Note that this apparent phase transition is only a finite volume effect. Our simulations are performed at zero temperature using symmetric $(L/a)^4$ volumes. The transition of the Polyakov loop indicates when the system size becomes comparable to the deconfinement length scale and transits from the small volume $\varepsilon$-regime to the large volume $p$-regime. In the limit of infinite volume, the pure Yang-Mills system is confining at all values of the bare gauge coupling.}
Since $\beta_b=5.30$ sits in the transition region, we discard it from our main analysis but use it to check for systematic effects later on.
We study the autocorrelation of the renormalized coupling $\gGF$ using the $\Gamma$-method \cite{Wolff:2003sm} and typically find integrated autocorrelation times of less than four measurements (80 MDTU). However, near the transition from the deconfined to the confined regime, the integrated autocorrelation times increase up to fifteen measurements (300 MDTU).

While simulations in the deconfined regime have zero topological charge, nonzero topological charges are expected in the confined region. As we decrease $\beta_b$ for a fixed volume, we indeed observe that nonzero topological charges arise and that their fluctuations increase. We study the effect of nonzero topological charge on $\gGF$ by ``filtering'' configurations according to topological sectors and compare the corresponding values of $\gGF$.
Our data suggest that for the fast-running pure gauge system, the $\beta$ function  is sufficiently large that the impact of nonzero topological charges is not statistically resolved.\footnote{The situation is quite different  for (near)-conformal simulations with many dynamical flavors, where the $\beta$ function runs slow and any topology   can  significantly  alter the final result \cite{Hasenfratz:2020vta}. }

\section{GF coupling and the RG \texorpdfstring{$\beta$}{beta} function}\label{sec:lat_g2}

\begin{figure*}[p]
    \centering
    \includegraphics[height=0.224\textheight]{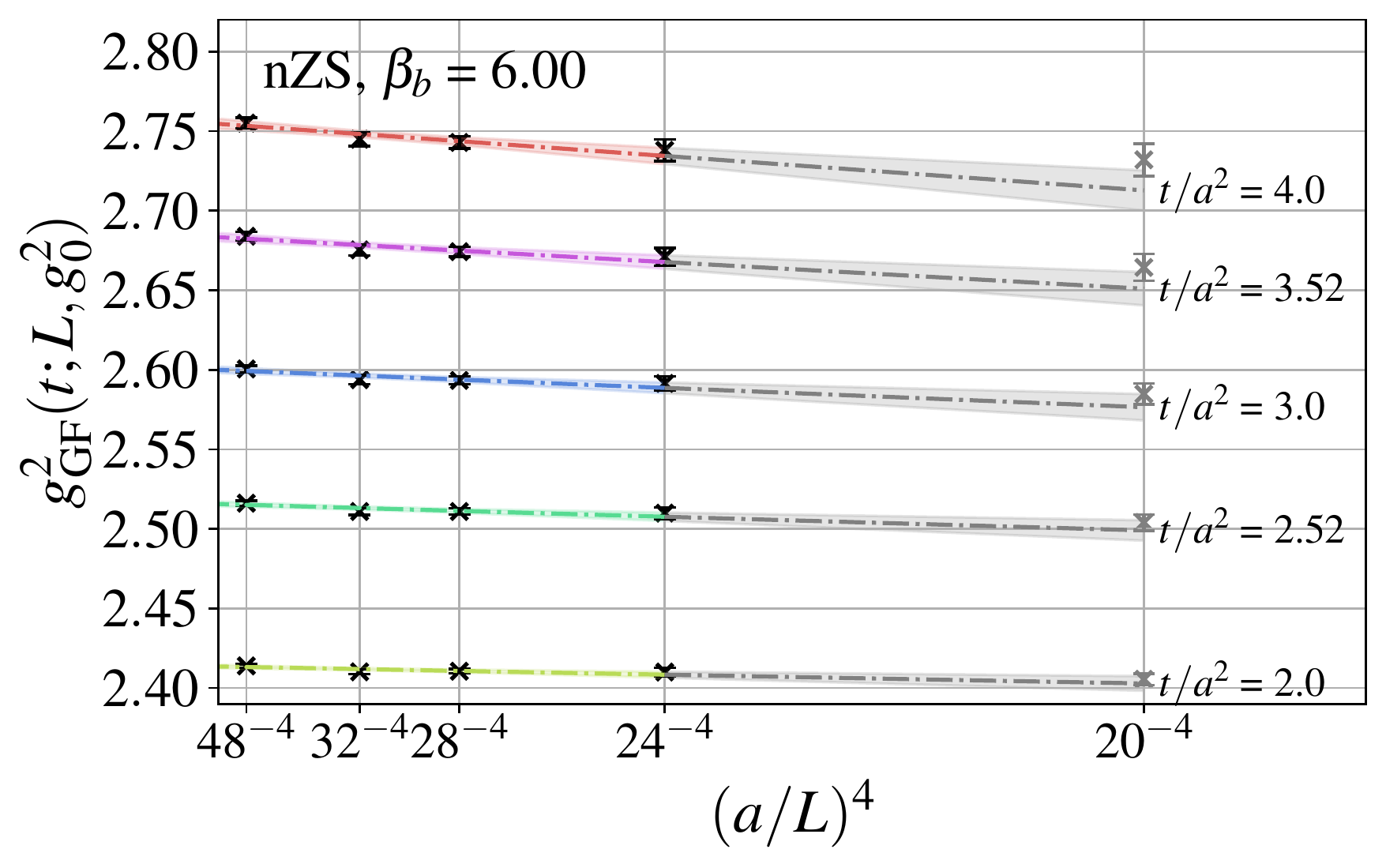}\hfill
    \includegraphics[height=0.224\textheight]{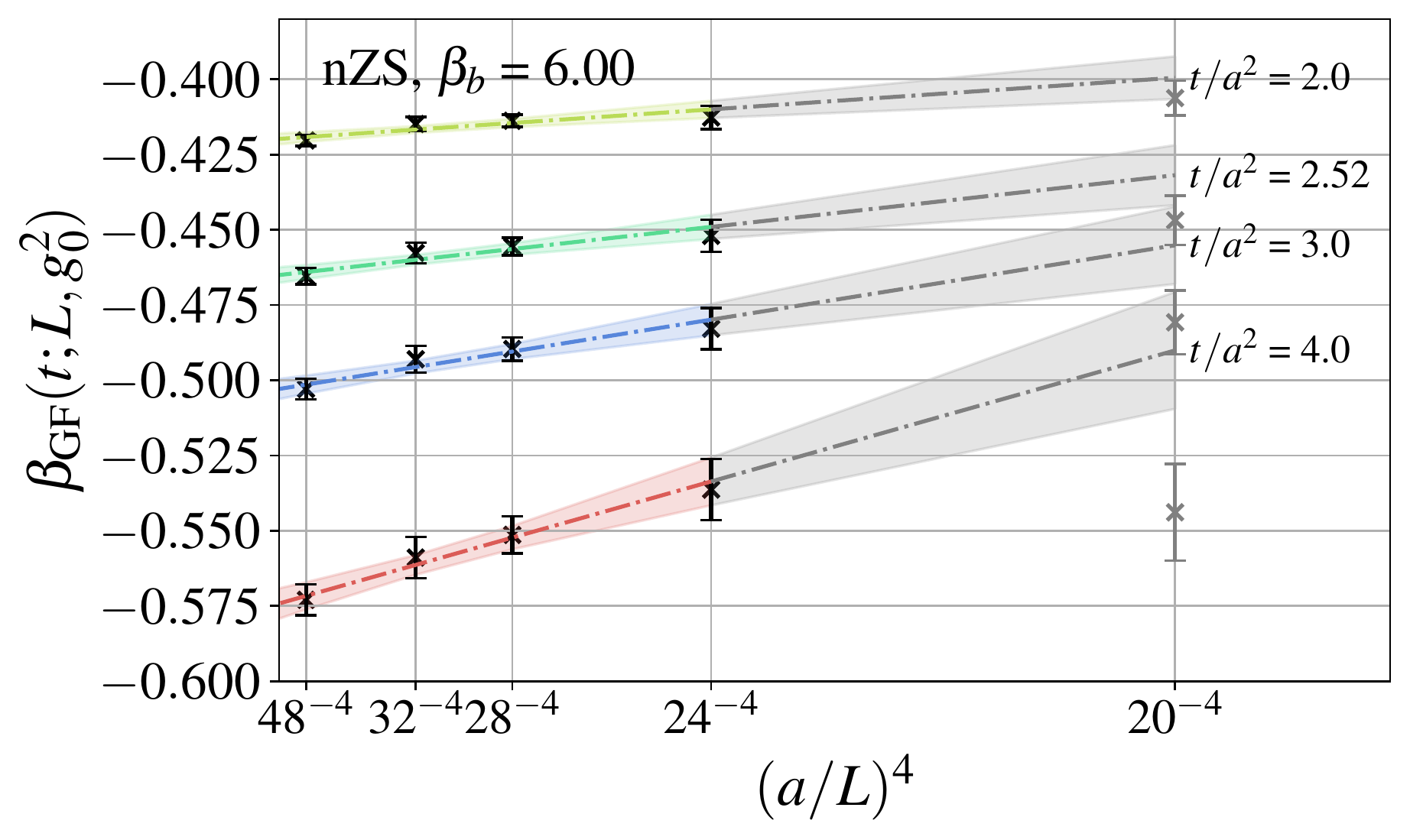}\\
    ~~~\includegraphics[height=0.224\textheight]{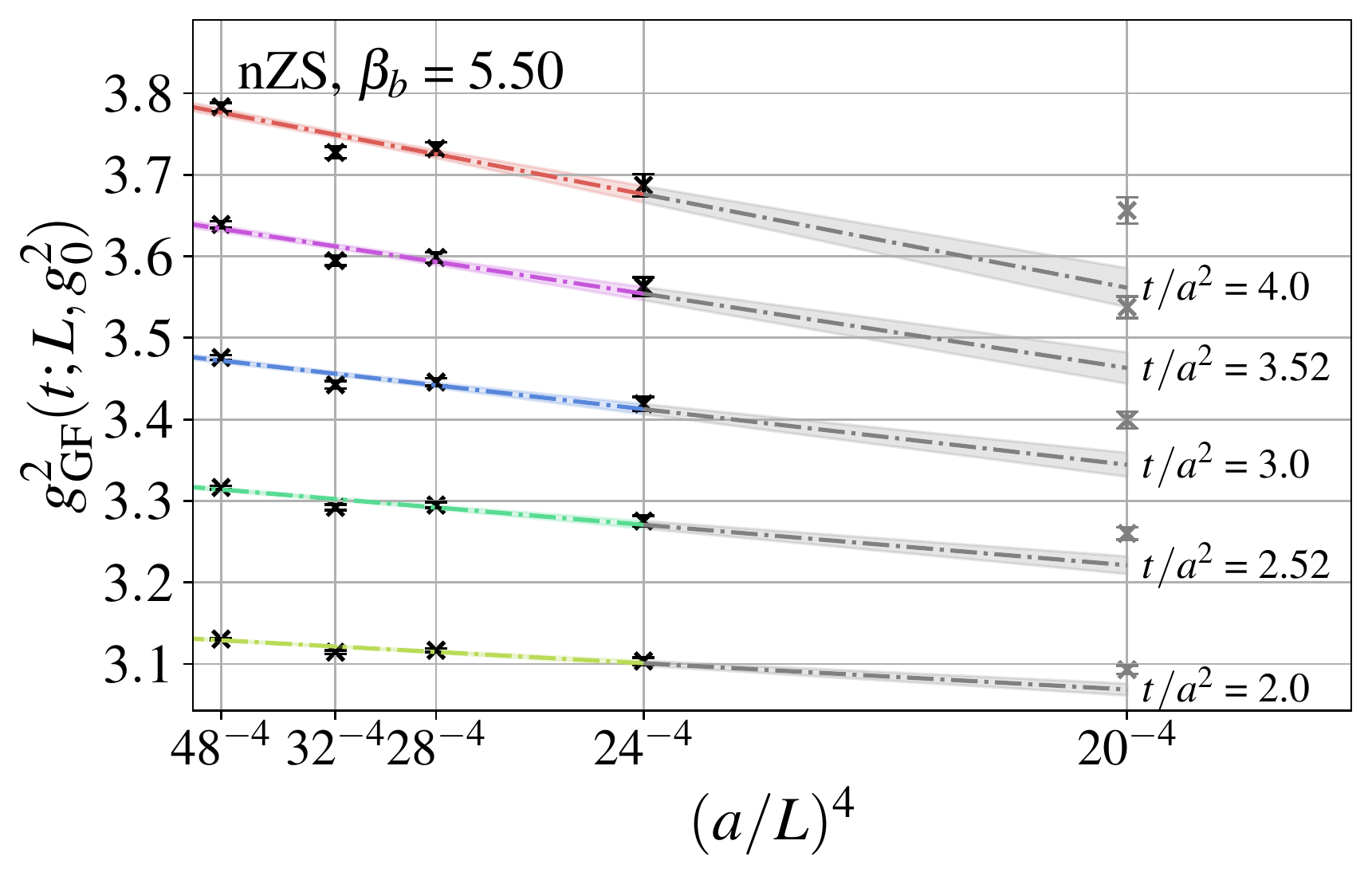}\hfill
    \includegraphics[height=0.224\textheight]{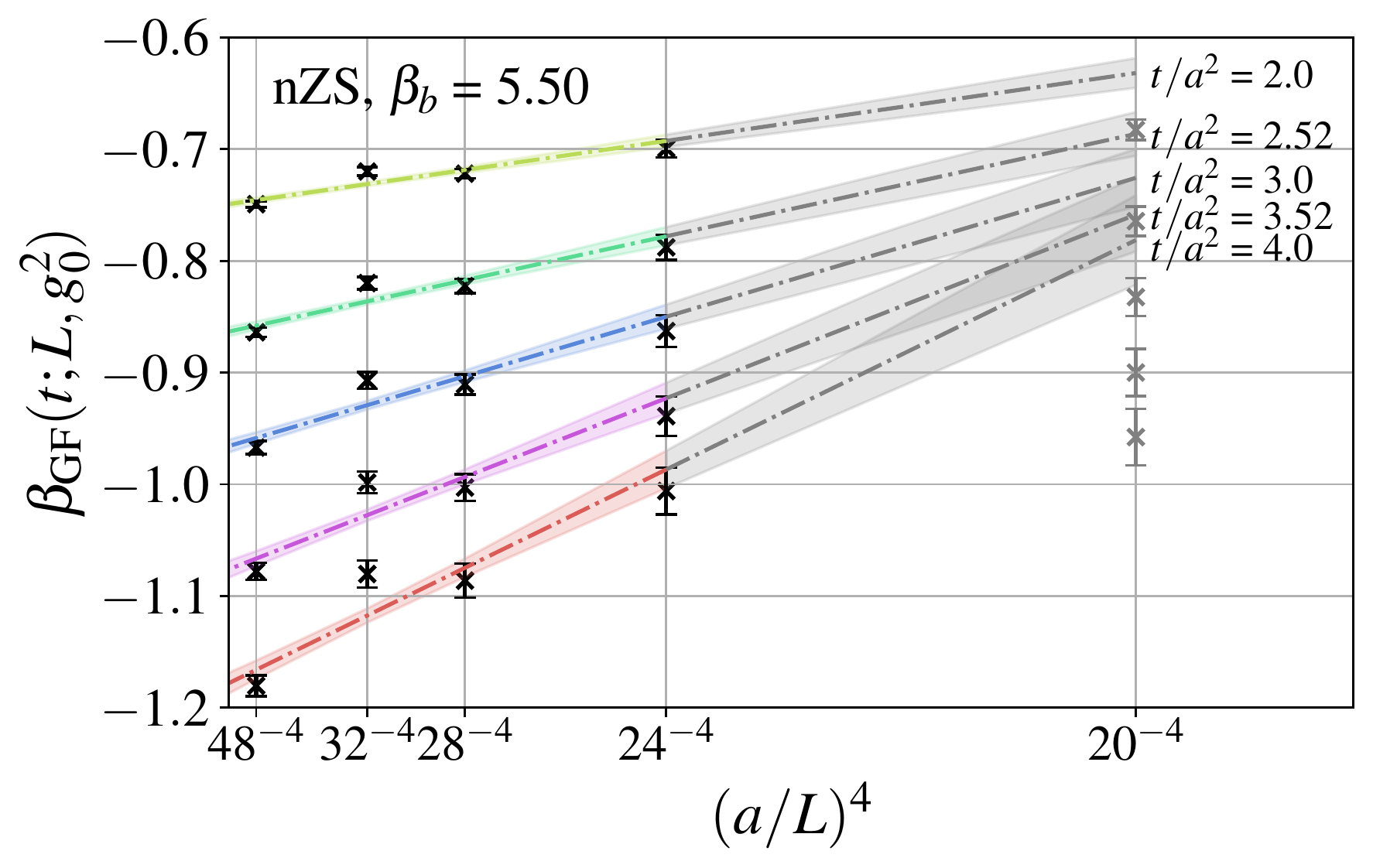}\\
    \includegraphics[height=0.224\textheight]{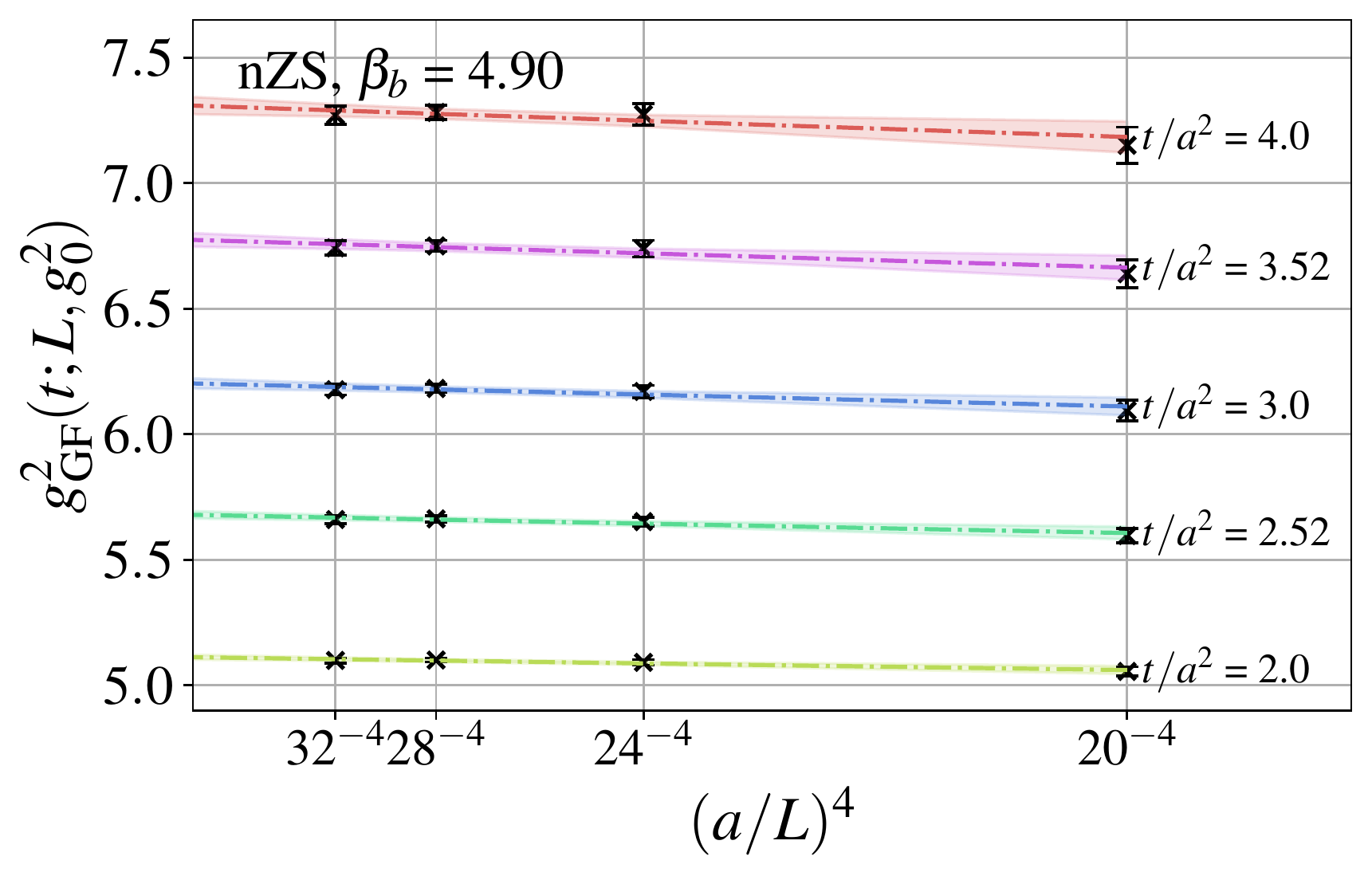}\hfill
    \includegraphics[height=0.224\textheight]{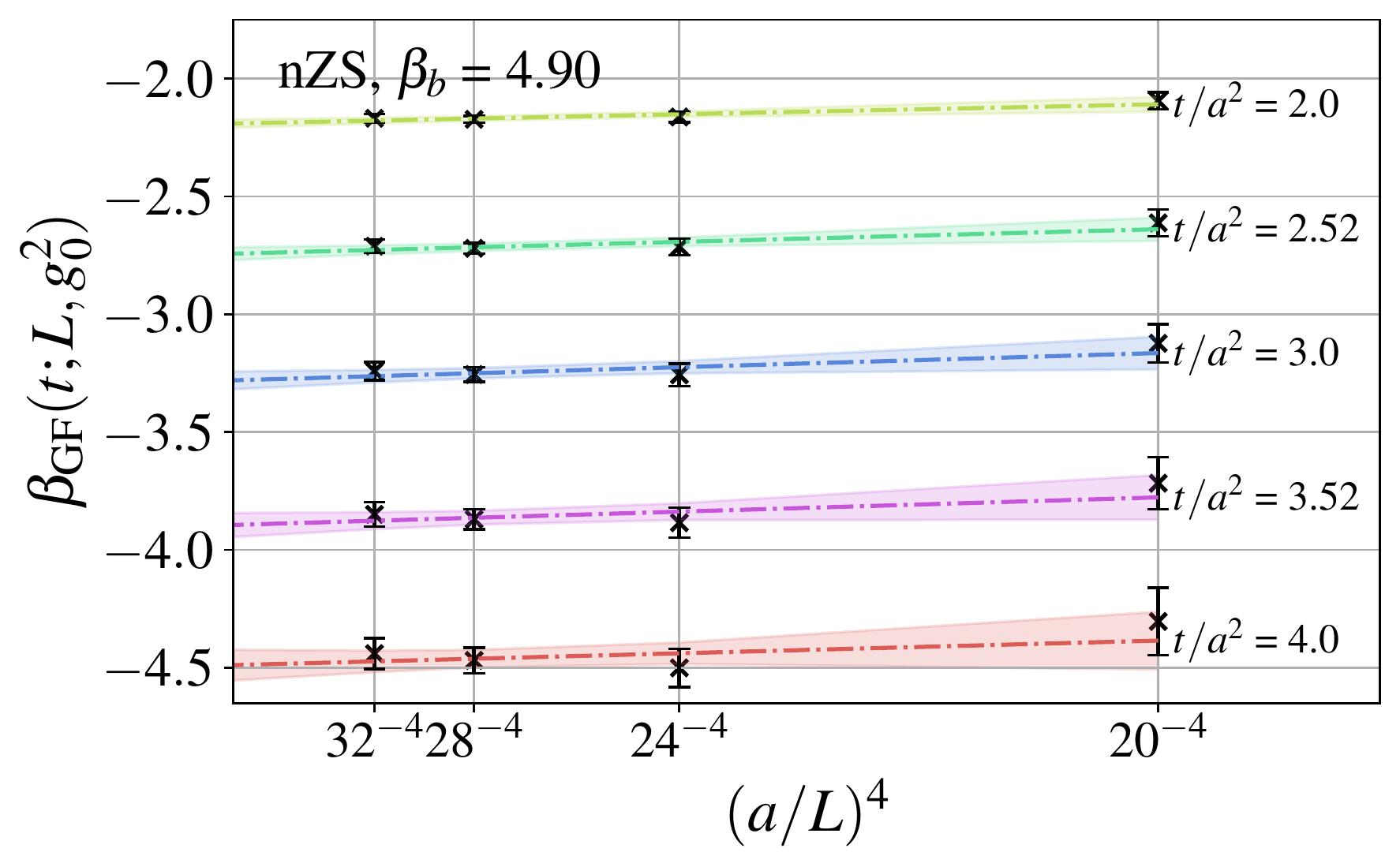}\\
    ~~~\includegraphics[height=0.224\textheight]{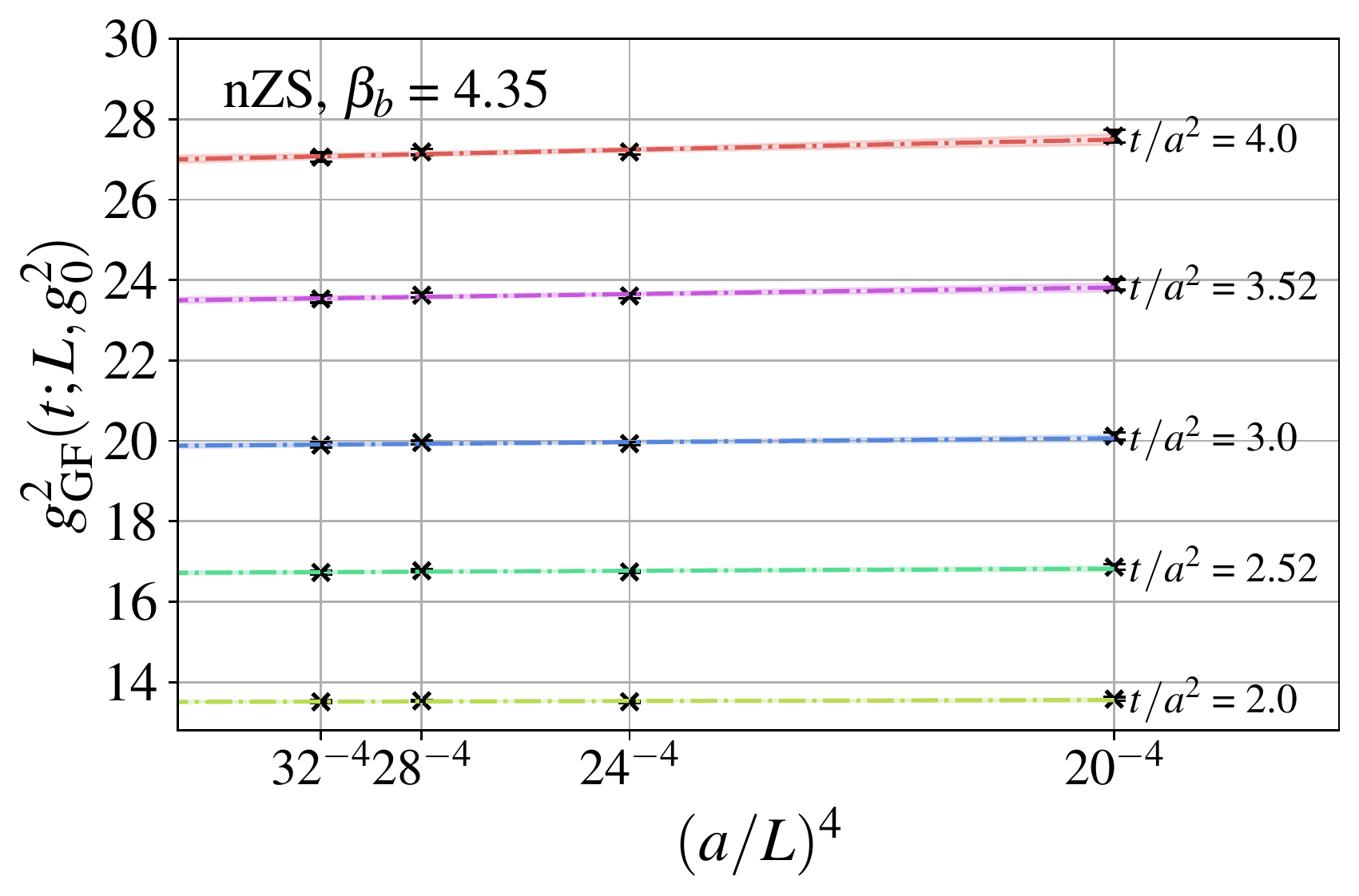}\hfill
    \includegraphics[height=0.224\textheight]{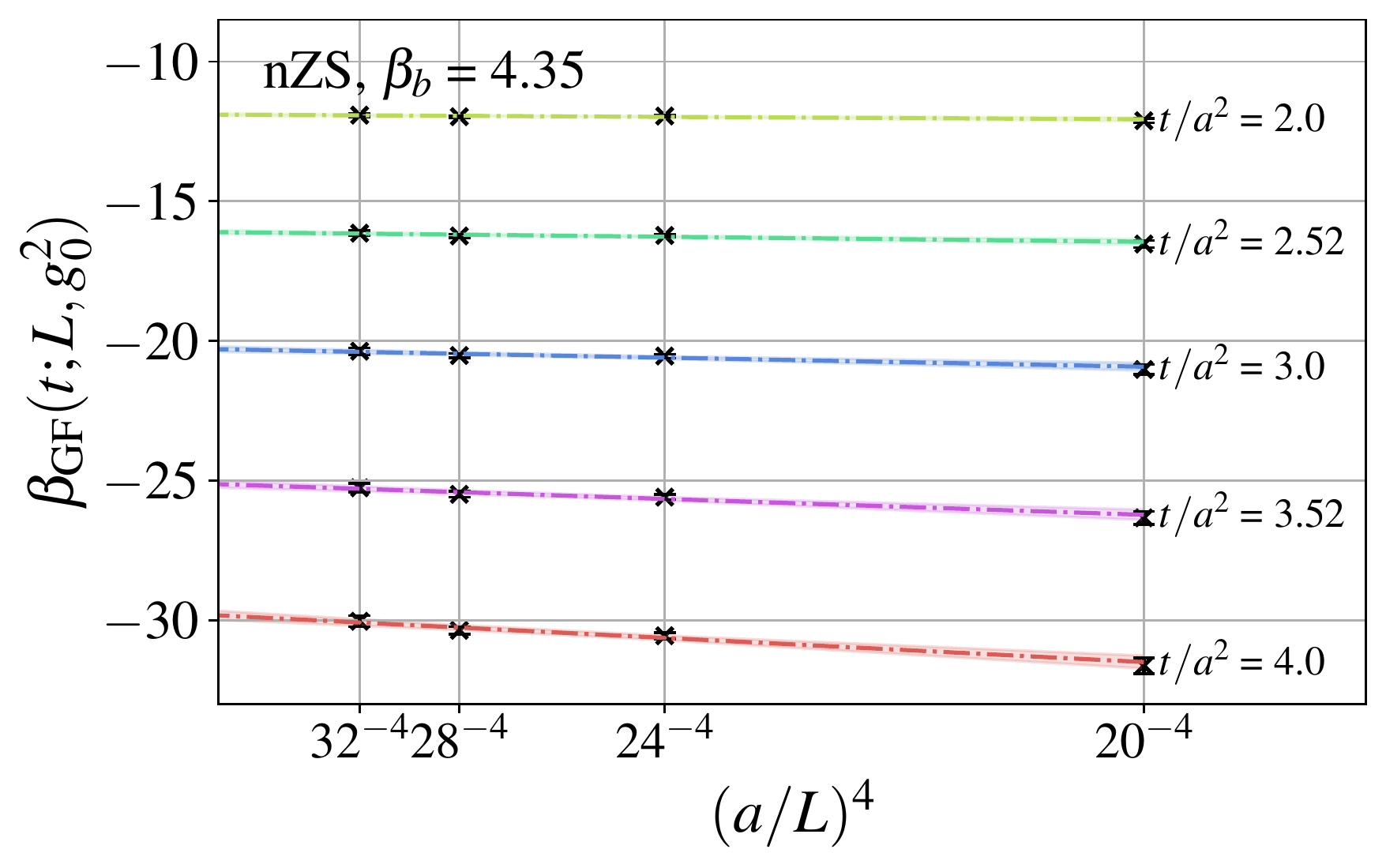}\\
    \caption{Examples of infinite volume extrapolation of $\gGF(t;L,g_0^2)$ (left) and $\betaGF(t;L,g_0^2)$ (right) at $\beta_b=6.00$, $5.50$, $4.90$ and  $4.35$ (top to bottom). Each color corresponds to a fixed lattice flow time between $t/a^2=2.0$ (yellow) to $t/a^2=4.0$ (red). Black  symbols indicate  data points that are included in the infinite volume extrapolation, whereas symbols shown in gray are not included and just shown for reference.}
\label{fig:inf_vol_extrap}
\end{figure*}

\begin{figure*}[tb]
    \centering
    \includegraphics[height=0.23\textheight]{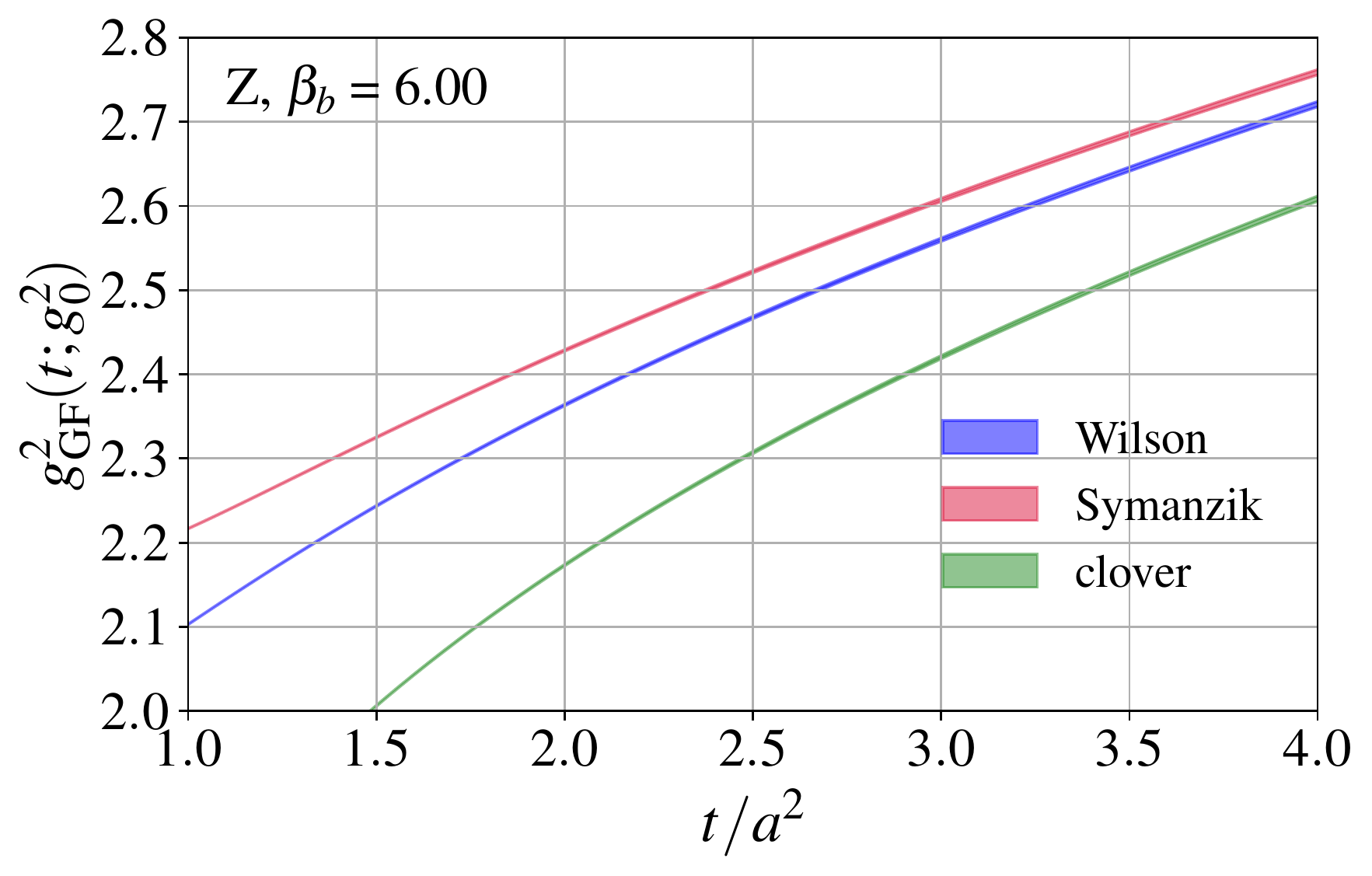}\hfill
    \includegraphics[height=0.23\textheight]{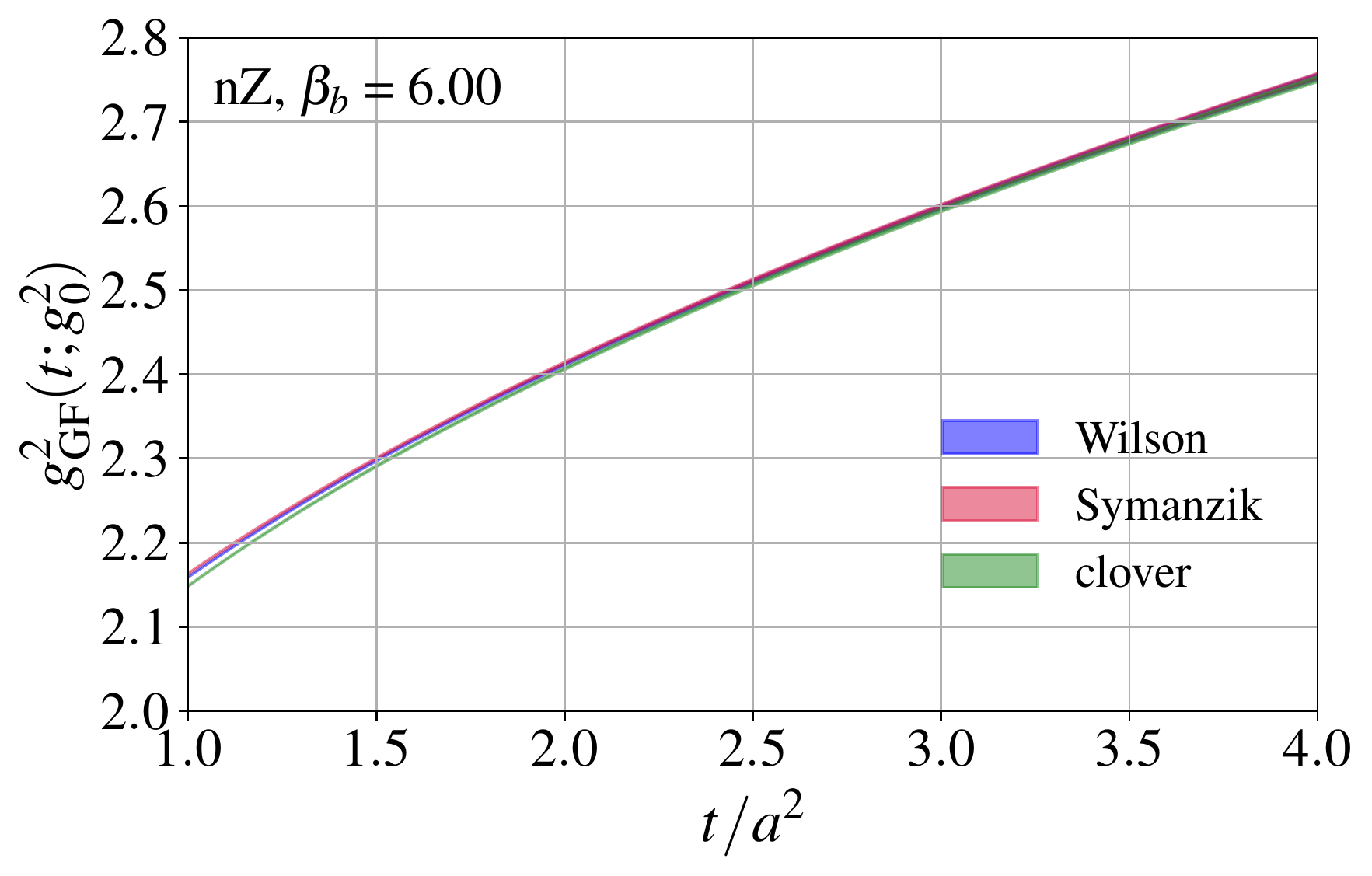}\\
    \includegraphics[height=0.23\textheight]{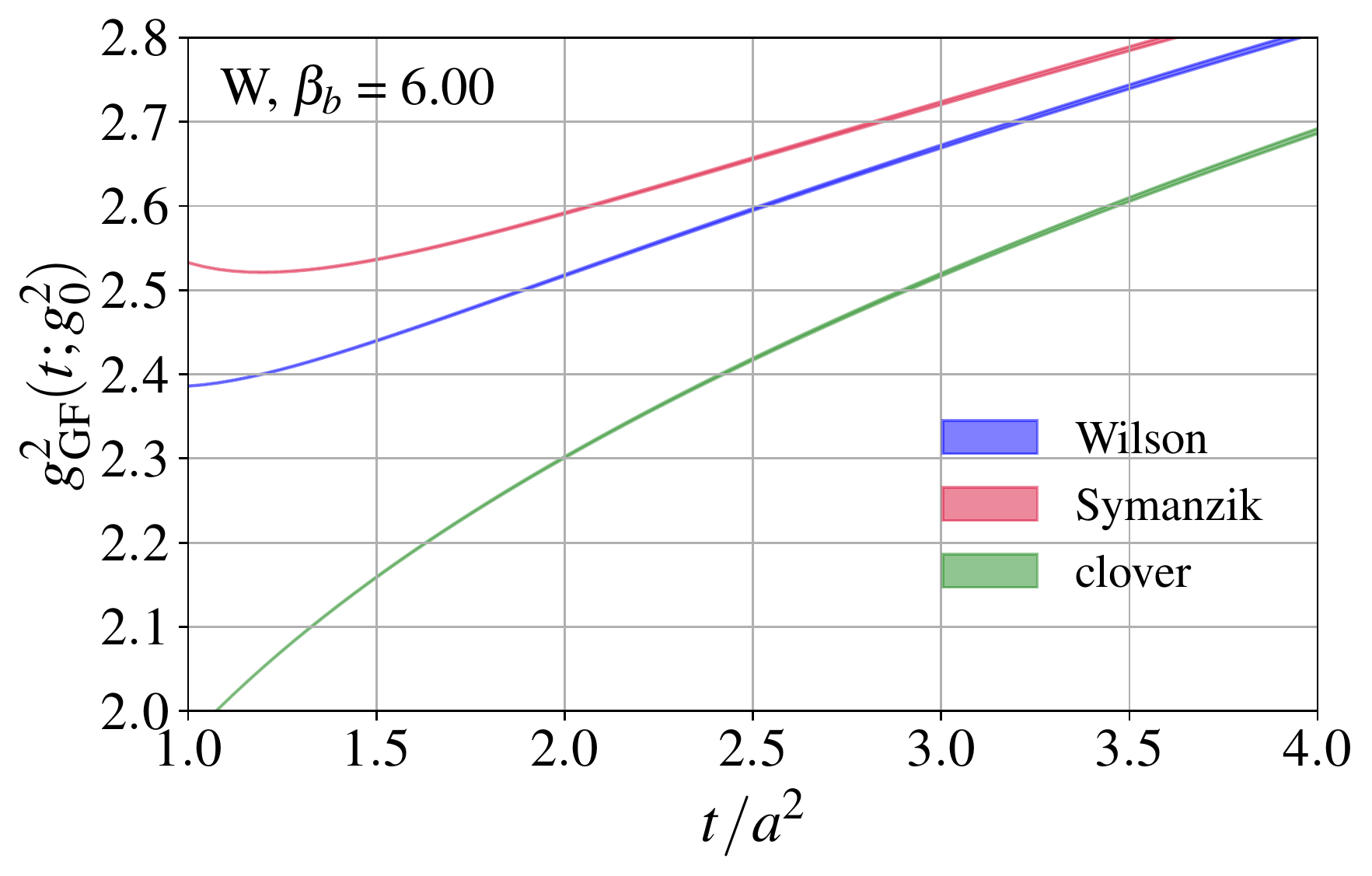}\hfill
    \includegraphics[height=0.23\textheight]{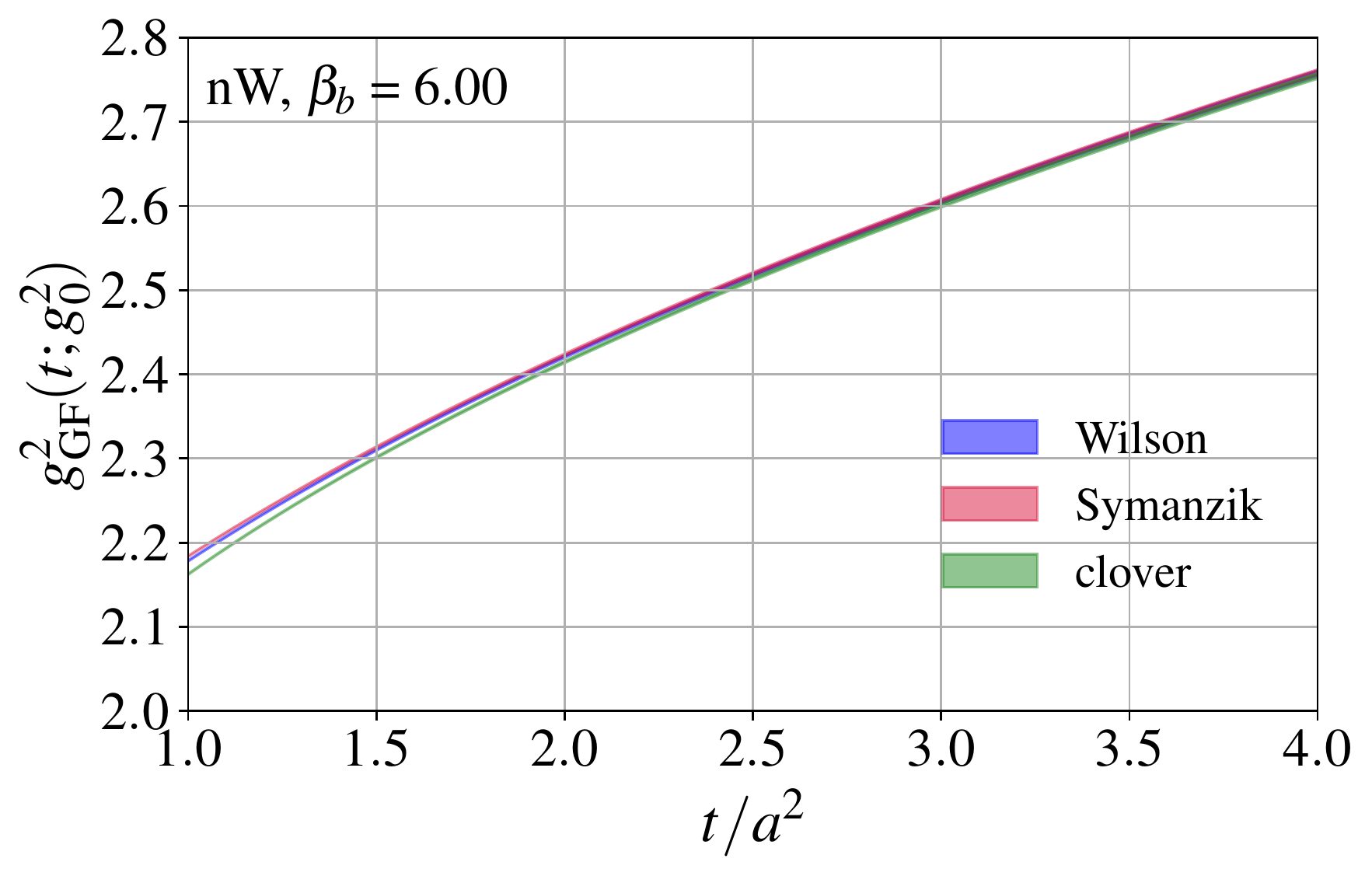}
    \caption{Infinite volume extrapolated gradient flow coupling  without tln (left) and with tln (right) as a function of $t/a^2$ at $\beta_b=6.0$. Different colors correspond to different operators. The energy density is  obtained for Zeuthen flow in the top and Wilson flow in the bottom panels.}
    \label{fig:tln_weak_comparison}
\end{figure*}

\begin{figure*}[tb]
    \centering
    \includegraphics[height=0.23\textheight]{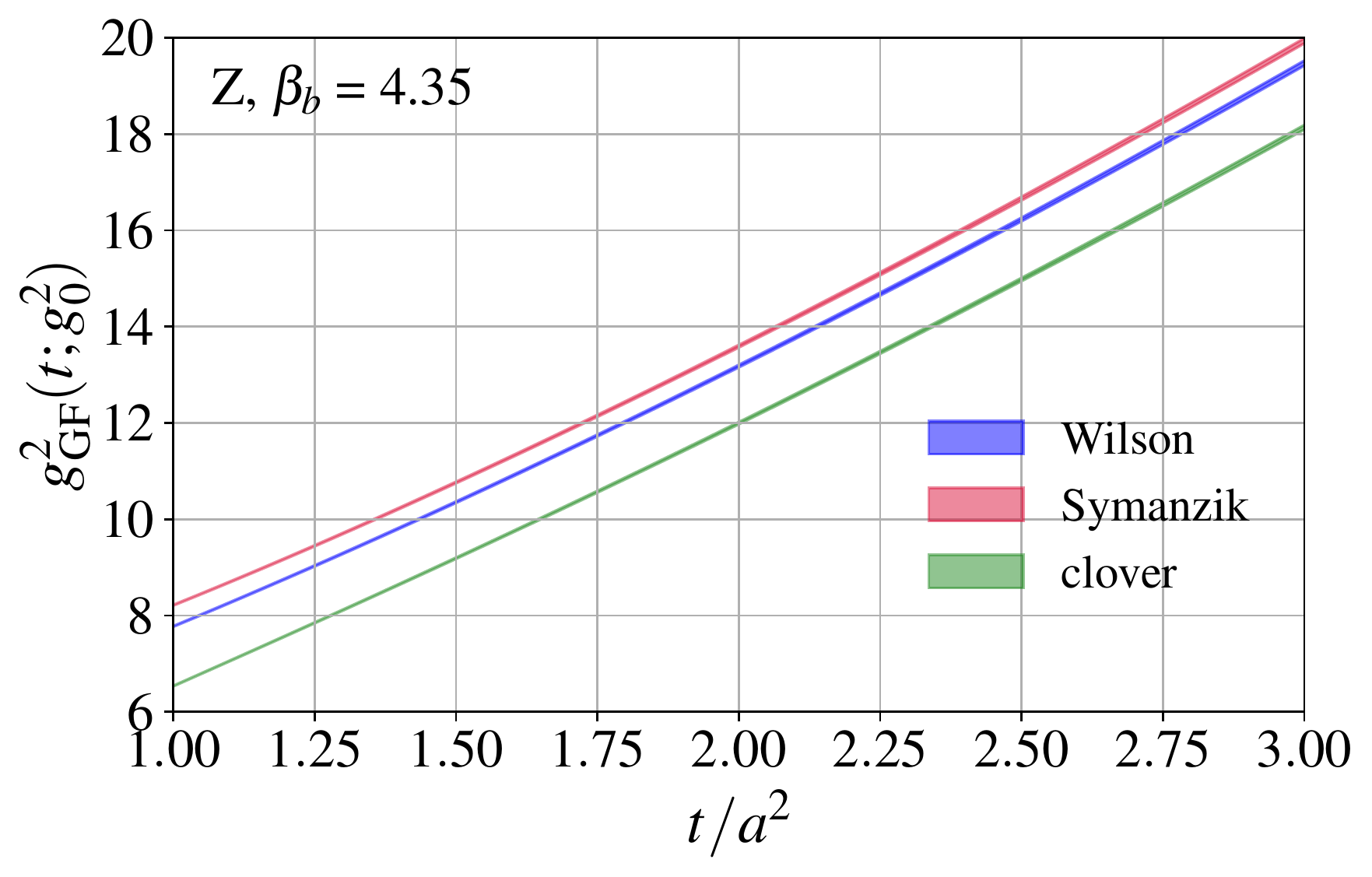}
    \includegraphics[height=0.23\textheight]{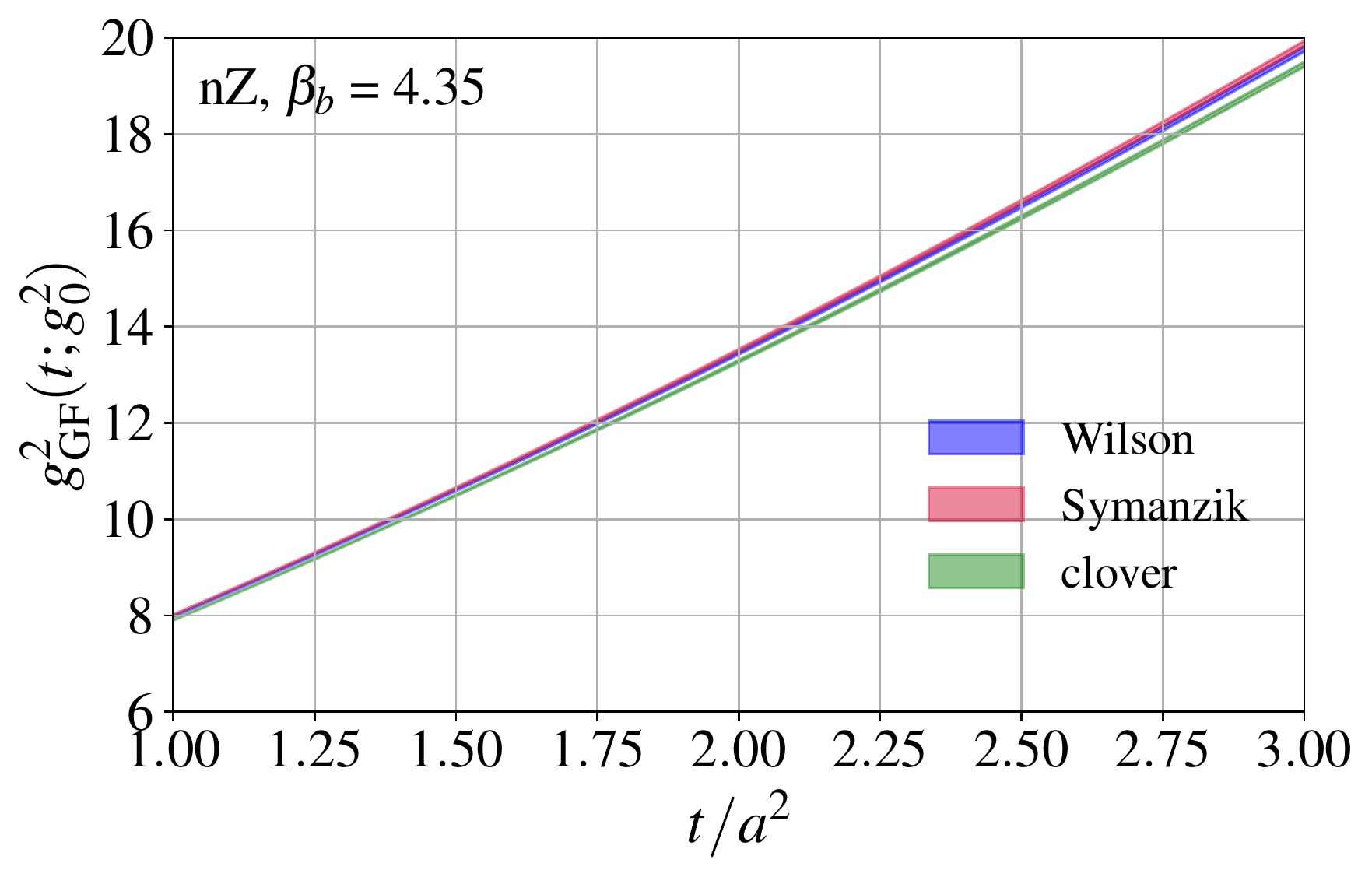}\\
    \includegraphics[height=0.23\textheight]{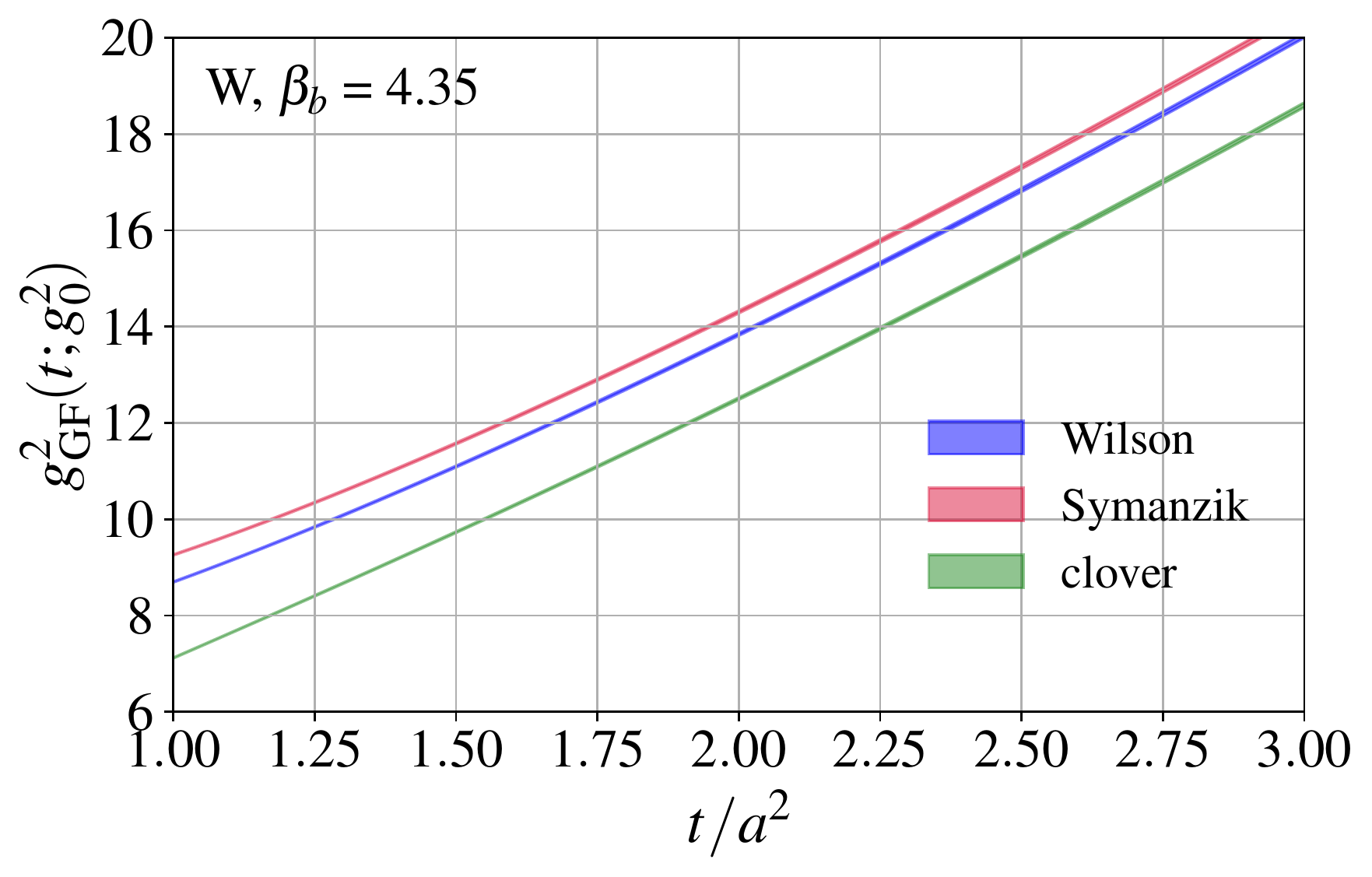}
    \includegraphics[height=0.23\textheight]{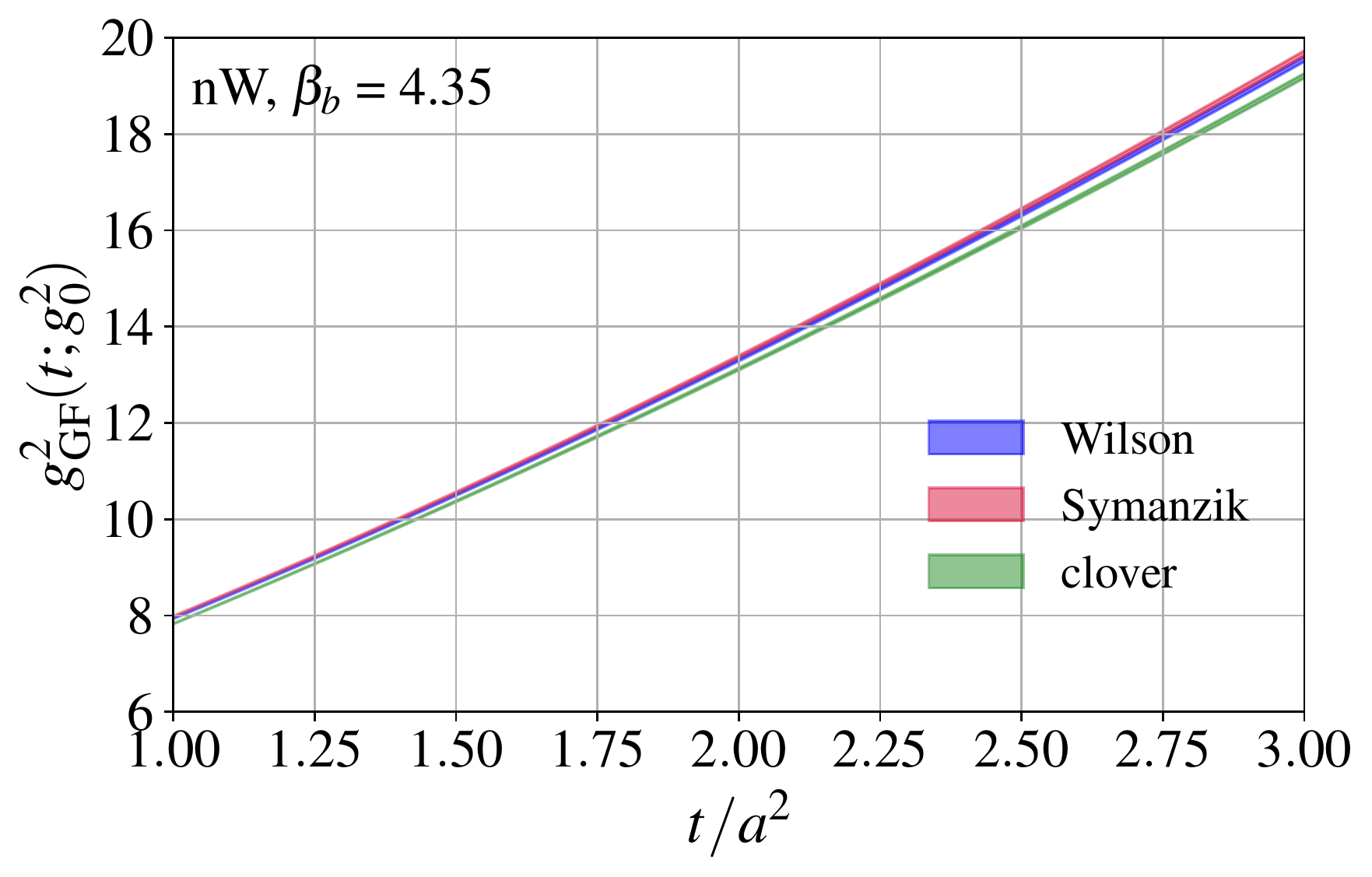}
    \caption{Infinite volume extrapolated gradient flow coupling  without tln (left) and with tln (right) as a function of $t/a^2$ at $\beta_b=4.35$. Different colors correspond to different operators. The energy density is obtained with Zeuthen flow in the top and Wilson flow in the bottom panels. }
    \label{fig:tln_strong_comparison}
\end{figure*}

\subsection{Definition of the GF coupling and \texorpdfstring{$\beta$}{beta} function}\label{sec:cpling_and_beta_functionmeas}

 We define the finite volume gradient flow coupling on the lattice  \cite{Luscher:2010iy,Fodor:2012td}  as
\begin{equation}\label{eqn:fv_GF_cpling}
\gGF\big(t;L,g_0^2\big) \equiv \frac{\mathcal{N}}{C(t,L)}\
\svev{t^2 E(t)},
\end{equation}
where $g^2_0=6/\beta_b$ is the bare gauge coupling. The term $C(t,L)=1+\delta(t,L)$ corrects for the  zero modes  caused by the periodic BC of the gauge fields \cite{Fodor:2012td}. The energy density $\svev{E(t)}$ can be approximated by local gauge observables.  In general, $\gGF(t;L,g^2_0)$ also depends on the gauge action, the gradient flow, and the operator chosen to approximate $\langle E(t) \rangle $.  In this work we use Symanzik improved gauge action  and consider two different gradient flow transformations, Wilson (W) and Zeuthen (Z) flow. We  determine the Wilson plaquette (W), the clover (C), and the tree-level improved Symanzik operator (S) to estimate $\langle E(t) \rangle $.  It is possible to calculate the tree-level lattice corrections to $\gGF(t;L,g_0^2)$ for a given action-flow-operator combination  and include those in $C(t,L)$ \cite{Fodor:2014cpa}.
In our analysis, we consider $C(t,L)$ with and without tree-level corrections, and we refer to the former as ``tree-level normalization'' (tln). We use  a shorthand notation  to distinguish the different flow-operator combinations, e.g.~ZS refers to Zeuthen flow and Symanzik operator. When using tln in the definition of $g^2_{\mathrm{GF}}(t;L,g_0^2)$ we  prepend  ``n'', i.e.~nZS refers to tln improved Zeuthen flow and Symanzik operator. In the continuum limit, the RG $\beta$ function is independent of the bare action, GF transformation or operator choice, and the comparison of the different combinations can serve to estimate systematic  effects. Continuum extrapolations on the lattice are more  stable when lattice artifacts are small, and we chose the nZS combination as our preferred analysis.

From the GF coupling in Eq.~(\ref{eqn:fv_GF_cpling}), we derive the GF $\beta$ function
\begin{equation}
  \betaGF\big(t;L,g_0^2\big) = -t\frac{\mathrm{d}}{\mathrm{d}t} \gGF\big(t;L,g_0^2\big)
  \label{Eq.betaGF}
\end{equation}
in finite volume by discretizing the flow-time derivative $\mathrm{d} / \mathrm{d}t$ with a 5-point stencil. The interval of the time derivative  is set by the time step $\epsilon$ used to integrate the gradient flow equation for the gauge field. We choose $\epsilon=0.04$ and explicitly verified that this choice has no impact on our analysis by repeating the GF measurements using $\epsilon=0.01$ for selected ensembles.

Our numerical analysis starts by determining the renormalized couplings $\gGF(t;L,g_0^2)$ at all flow times for our three operators. We determine $\gGF(t;L,g_0^2)$ for both data sets obtained with Zeuthen and Wilson flow, respectively, and for our entire set of  gauge field ensembles. Next, we  numerically calculate the derivative as specified in Eq.~(\ref{Eq.betaGF}). The uncertainties are propagated using standard correlated error propagation techniques  implemented in the software packages \texttt{gvar} \cite{gvar} and \texttt{lsqfit} \cite{lsqfit}.

The analysis of the RG $\beta$ function proceeds by first taking the infinite volume limit $a/L \rightarrow 0$ of both $\gGF(t;L,g_0^2)$ and $\betaGF(t;L,g_0^2)$ independently at fixed $\beta_b$ and $t/a^2$. This is followed by an interpolation of $\betaGF(t;g_0^2)$ in $\gGF(t;g_0^2)$  at fixed  $t/a^2$. The last step of our analysis is to take the $a^2/t\to 0$ continuum limit at fixed $\gGF$. These steps are detailed in the rest of this section.
\subsection{Infinite volume extrapolation}\label{subsec:analysis_methods}

In a 4-dimensional gauge-fermion system the volume is a relevant parameter, and the RG equation in finite volume includes a term describing its effect on the running of the renormalized coupling. We prefer to avoid complications arising from this a priori unknown quantity and define the RG $\beta$ function in the infinite volume limit.  This requires first extrapolating to infinite volume before considering the continuum limit.

Since the energy density $\svev{E(t)}$ is a dimension-4 operator, the finite volume corrections of $\gGF$ are expected to be $\mathcal{O}((a/L)^4)$ at leading order. We independently extrapolate both  the GF coupling $\gGF(t;L,g_0^2)$ and GF $\beta$ function $\betaGF \big(t;L,g_0^2\big)$ linearly in  $(a/L)^4 \rightarrow 0$ for each fixed bare gauge coupling $\beta_b$ and lattice flow time $t/a^2$. This  analysis strategy was first outlined in Ref.~\cite{Peterson:2021lvb}. Alternative methods are discussed, e.g., in Refs.~\cite{Fodor:2017die, Hasenfratz:2019puu, Hasenfratz:2019hpg, Kuti:2022ldb}.

In Fig.~\ref{fig:inf_vol_extrap}  we show typical infinite volume extrapolations for $\gGF(t;L,g_0^2)$ and $\betaGF(t;L,g_0^2)$ at relatively weak coupling ($\beta_b=6.0$), at intermediate couplings ($\beta_b=4.9$ and 5.50), and in the strong coupling regime ($\beta_b=4.35$). Each panel shows the $(a/L)^4$ extrapolation at five different lattice flow time values that cover the range we use in the continuum limit extrapolation (cf.~Sec.~\ref{subsec:cont_lim}).  In the strong coupling regime, we observe very mild volume dependence. This is consistent with the expectation that below the confinement scale $\mu_\text{conf}$, i.e.~at renormalized couplings stronger than $\gGF(\mu_\text{conf})$,  the confinement scale provides an infrared cutoff and the volume dependence is suppressed. 
Thus we find it sufficient to use volumes with  $L/a=\{20,\, 24,\, 28,\, 32\}$ in this regime. At renormalized couplings that correspond to energy scales above the confinement scale, the volume dependence is more significant,  as the only infrared cutoff is due to the finite volume. Here we drop the $L/a=20$ ensembles and add $48^4$ lattices  in the infinite volume extrapolation. At $\beta_b=5.5$ and 6.0  volumes $L/a \leq 32$ appear deconfined, while $L/a=48$ is transitioning to deconfined at $\beta_b=6.0$ and is confined at $\beta_b=5.5$. Similarly the $L/a=32$ at $\beta=5.5$ ensemble is transitioning and exhibits very long autocorrelation times. Likely these long autocorrelations result in underestimated statistical errors causing the $2\sigma$ deviation. 
At $\beta_b=5.0$ the Polyakov loop expectation value shows that volumes $L/a\ge 28$ are confining, while $L/a=20$ and 24 are in the transition region. $L/a=24$ shows especially large autocorrelation times. Nevertheless, at $\beta_b=5.0$ we use all five volumes in the infinite volume extrapolation, though it is dominated by the larger volumes.   We always perform  the infinite volume extrapolation using a linear fit ansatz in $(a/L)^4$. For flow times $2.0 \lesssim t/a^2 \lesssim 4.0$ contributing to the continuum extrapolations in Sec.~\ref{subsec:cont_lim}, most fits have good $p$-values ($\gtrsim 10\%$).  Notable exceptions are $\beta_b=5.0$ and 5.5, where large autocorrelations make it difficult to estimate the errors correctly. When adding $\beta_b=5.3$ to our analysis,  the infinite volume extrapolations have   vanishing $p$-values. The situation did not improve despite extending the Monte Carlo simulations and using larger thermalization cuts. We therefore conclude that $\beta_b=5.3$ suffers from sitting in the transition region and discard it from our main analysis.

\subsection{Different flows, operators and tree-level improvement }\label{sec:improvement}

In Figs.~\ref{fig:tln_weak_comparison}  and \ref{fig:tln_strong_comparison} we show the infinite volume extrapolated GF coupling $\gGF(t;g_0^2)$ as a function of the  gradient flow time $t/a^2$ with  and without tln improvement at $\beta_b=6.0$ and 4.35, respectively. We consider both Zeuthen and Wilson flow and all three operators.  In the continuum limit the different flows and operators must agree, so any difference between them at finite bare coupling points to cutoff effects. Figure \ref{fig:tln_weak_comparison} showcases the improvement due to  tln at weak coupling. As the left panels  show, there  are significant differences between the three operators even at flow time $t/a^2=4.0$. Comparing the top and bottom panels on the left, large differences for $\gGF(t; g_0^2)$ between Zeuthen (top) and Wilson (bottom) flow can be seen.  Both the operator and flow dependence are largely removed by tln normalization, which is demonstrated in the panels on the right.

Figure \ref{fig:tln_strong_comparison}  shows the same quantities at a strong bare coupling, $\beta_b=4.35$. The gauge coupling runs fast  and the same flow time range covers nearly 20 times the range of $g^2_{\mathrm{GF}}$ at $\beta_b=4.35$  compared to $\beta_b=6.0$. While the different operators and flows without tln correction may appear to be closer at $\beta_b=4.35$ than at $\beta_b=6.0$, the absolute  difference is much larger. Tree-level correction works similarly in the strong coupling, removing most flow and operator dependence. The somewhat larger spread after tln suggests that the perturbatively motivated improvement is not as effective in the strong coupling regime as at weak coupling.
The effect of tln improvement at strong coupling was recently studied in the context of gradient flow scale setting for the case of 2+1 flavor gauge field configurations generated with Iwasaki gauge action and Shamir domain-wall fermions \cite{Schneider:2022wmb}. That work also showed that tln removed most flow and operator dependence, but the improved predictions still had significant cutoff effects.
However, here we study the pure-gauge system where no effect due to fermion loops can spoil the improvement and, furthermore, we use the perturbatively improved Symanzik gauge action in comparison to Iwasaki gauge action.
 In Sec.~\ref{subsec:cont_lim}  we will show that in our case tln improvement not only reduces the difference between different flows and operators but also reduces cutoff effects.

 Based on the improvement from tln, we designate the tree-level improved Zeuthen flow with Symanzik operator  combination (nZS) as our preferred analysis. In the remainder of this paper we  therefore concentrate on analyzing this combination. Other tln improved flow-operator combinations are consistent within uncertainties and considered as part of our discussion on systematic uncertainties in \ref{subsec:sysbetafn}. Moreover, we repeat our analysis without tln improvement, which increases discretization errors. Within these more sizeable uncertainties, the unimproved combinations are consistent with our preferred nZS prediction.

\begin{figure}[tb]
    \centering
    \includegraphics[width=\columnwidth]{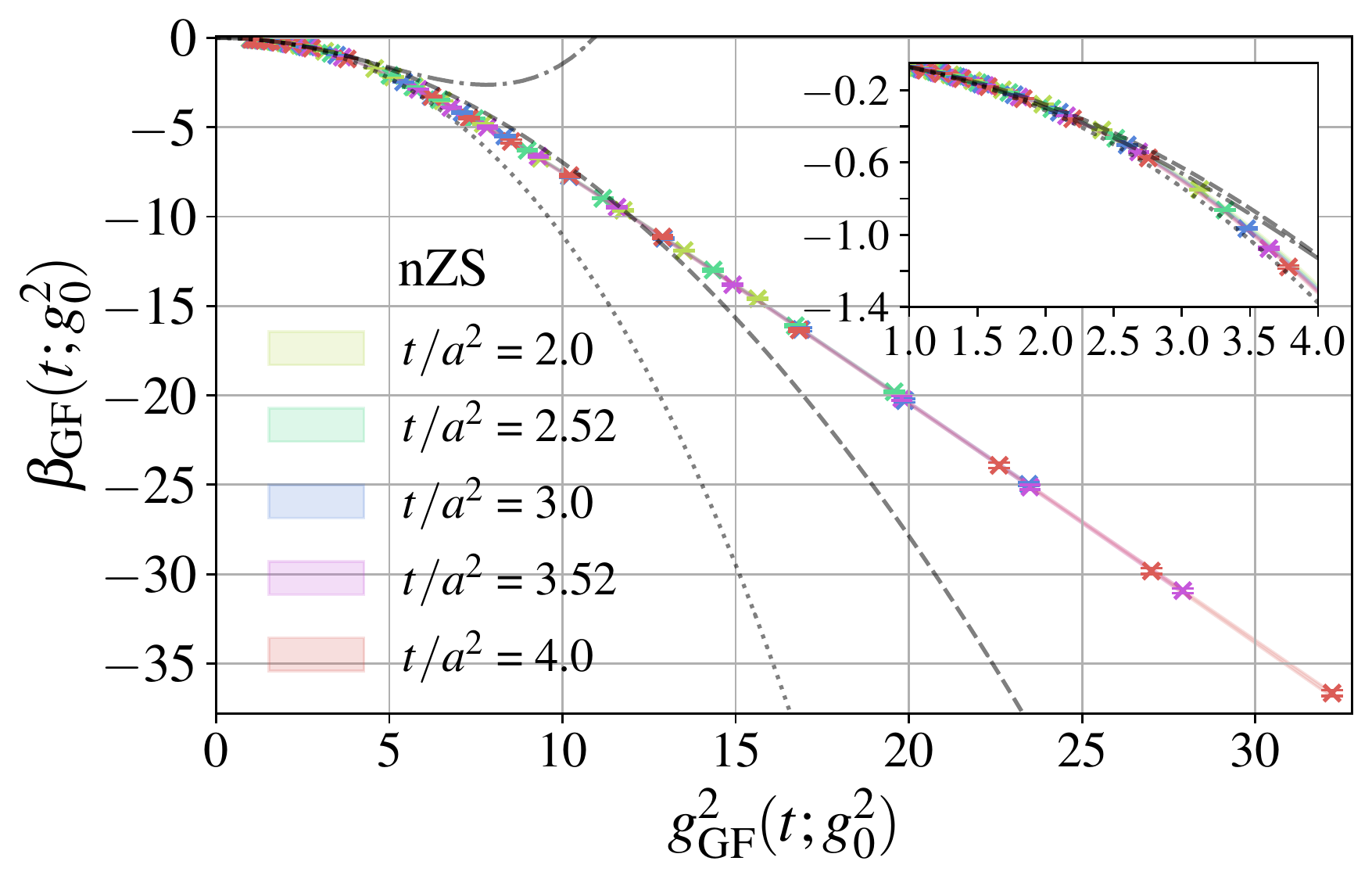}\\
    \includegraphics[width=\columnwidth]{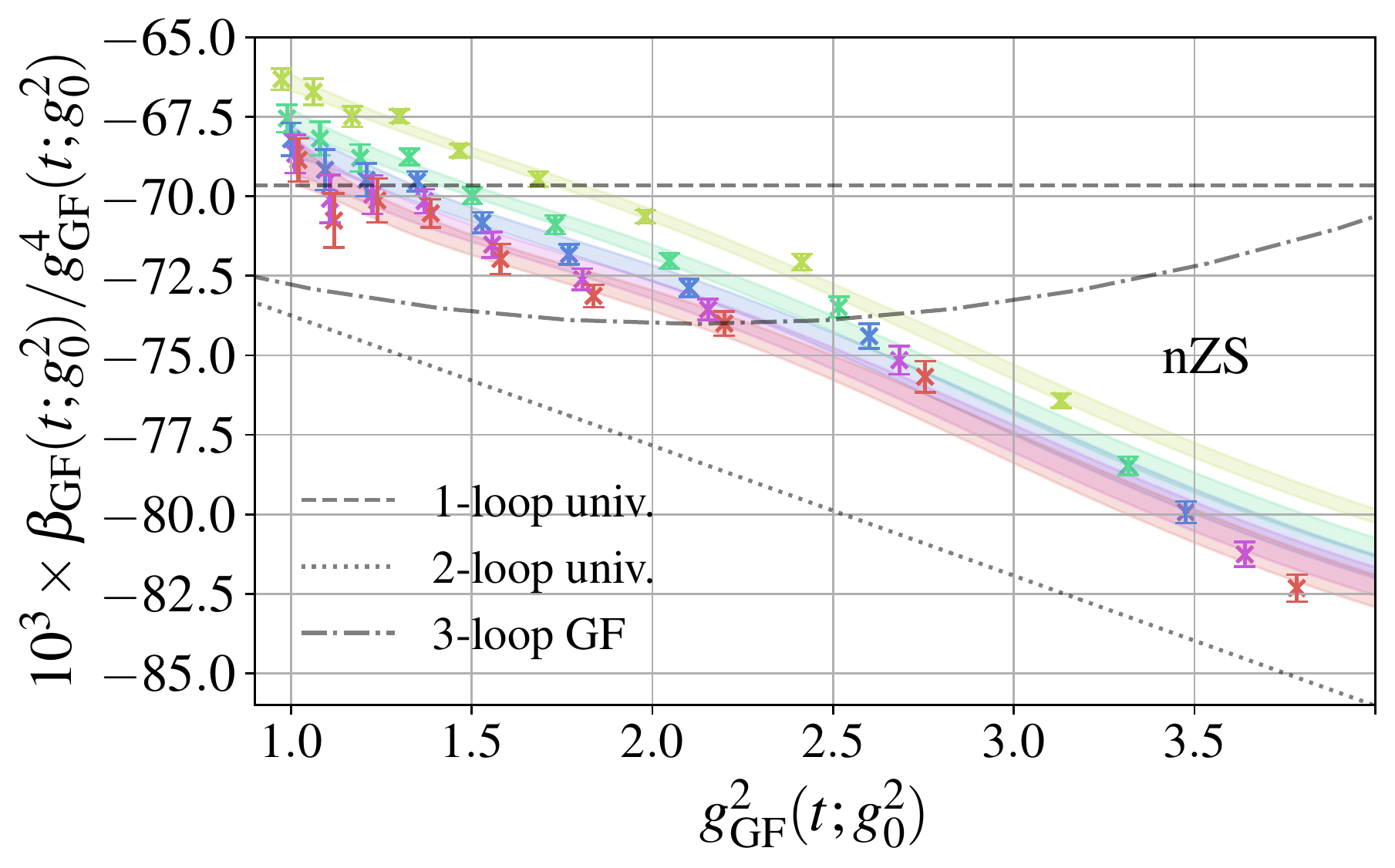}
    \caption{Interpolation of $\betaGF$ in terms of $\gGF$ at fixed flow times. Different colored symbols correspond to  ($\gGF$, $\betaGF$) pairs at flow times $t/a^2=2.0$ -- 4.0. The colored bands are the interpolating functions based on ratios of polynomials given in Eq.~(\ref{eq:pade}). In the lower panel, we plot $\betaGF(\gGF)/g^4_{\mathrm{GF}}$ to enhance the weak coupling regime.}
    \label{fig:interpolation}
\end{figure}

\subsection{Interpolation of \texorpdfstring{$\betaGF(\gGF)$}{betaGF} vs.~\texorpdfstring{$\gGF$}{gGF}}\label{subsec:interpolation}

The last task before we can take the $a^2/t \to 0$ continuum limit is the identification of $\{\betaGF\big(t;g^2_0),\,t/a^2\} $ pairs at fixed $\gGF\big(t;g^2_0)$. We achieve this by interpolating $\betaGF(t;g^2_0)$ in terms of $\gGF(t;g^2_0)$ at fixed values of lattice flow times $t/a^2$.

Our interpolating form has to be able to describe the weak coupling regime where we expect perturbative behavior $\beta(g_{\mathrm{GF}}^2)\sim -b_0 g_{\mathrm{GF}}^4 + \mathcal{O}\big(g_{\mathrm{GF}}^{6}\big)$  as well as the strong coupling regime where it appears that $\beta_{\mathrm{GF}}(g_{\mathrm{GF}}^2)\approx c_0 + c_1 g_{\mathrm{GF}}^2$. A polynomial is not appropriate to cover both  regimes. Instead, we choose a form  that is similar to the ratio of polynomials used in  Pad\'e approximations 
\begin{align}
\mathcal{I}_{N}\big(g_{\mathrm{GF}}^2\big) \equiv \frac{-p_{0}g_{\mathrm{GF}}^{4}\Big(1+\sum_{i=1}^{N}p_{i}g_{\mathrm{GF}}^{2i}\Big)}{     1+\sum_{j=1}^{N+1}q_{j}g_{\mathrm{GF}}^{2j}},
\label{eq:pade}
\end{align}
such that we enforce  the leading order $\betaGF(\gGF)\sim g_{\mathrm{GF}}^4$ behavior. Since cutoff effects can change even the asymptotic behavior of the $\beta$ function, this  ansatz could constrain the lower limit on the flow time $t/a^2$ used in the analysis. However with $N=4$  we find that $\mathcal{I}_{N}\big(\gGF\big)$  provides a good description with $p$-values between 17\% -- 32\% for  flow times $2.0 \lesssim t/a^2 \lesssim 4.0$.

The two panels of Fig.~\ref{fig:interpolation}  illustrate the interpolation at  various flow times in the range $t/a^2\in[2.0, 4.0]$. The colored data points on both panels correspond to infinite volume limit $(\gGF,\,\betaGF)$  pairs  at each  bare coupling and five selected flow time values. For reference we show the universal 1- and 2-loop and the GF 3-loop \cite{Harlander:2016vzb} perturbative  lines in gray using dashed, dotted, and dash-dotted lines, respectively. The colored bands describe the interpolation according to Eq.~(\ref{eq:pade}) with $N=4$.
The  top panel of Fig.~\ref{fig:interpolation} shows $\betaGF$ vs.~$\gGF$ in the entire range of our data. The data points at different bare couplings and flow times form a smooth curve, indicating that cutoff effects are mild. The corresponding interpolating curves basically overlap, creating the purple-hued band. This plot also illustrates that the $\beta$ function is approximately linear in the strong coupling regime. We will discuss this feature further in Sec.~\ref{subsec:strong_coup_reg}.  In the  insert   of the top panel  we zoom in on the weak coupling regime.
To enhance the small $\gGF$ behavior,  we plot $\betaGF(\gGF)/g_{\mathrm{GF}}^4$ vs.~$\gGF$ on the  bottom panel. At large flow times and small couplings, $\betaGF(\gGF)$ has to be consistent with the perturbative predictions. As the panel shows, both  the data and the interpolating forms are close to the perturbative values in our selected flow time range, though we do not  constraint the  intercept of $\betaGF(\gGF)/g^4_{\textrm{GF}}$.

\begin{figure}[tb]
    \centering
    \includegraphics[width=0.95\columnwidth]{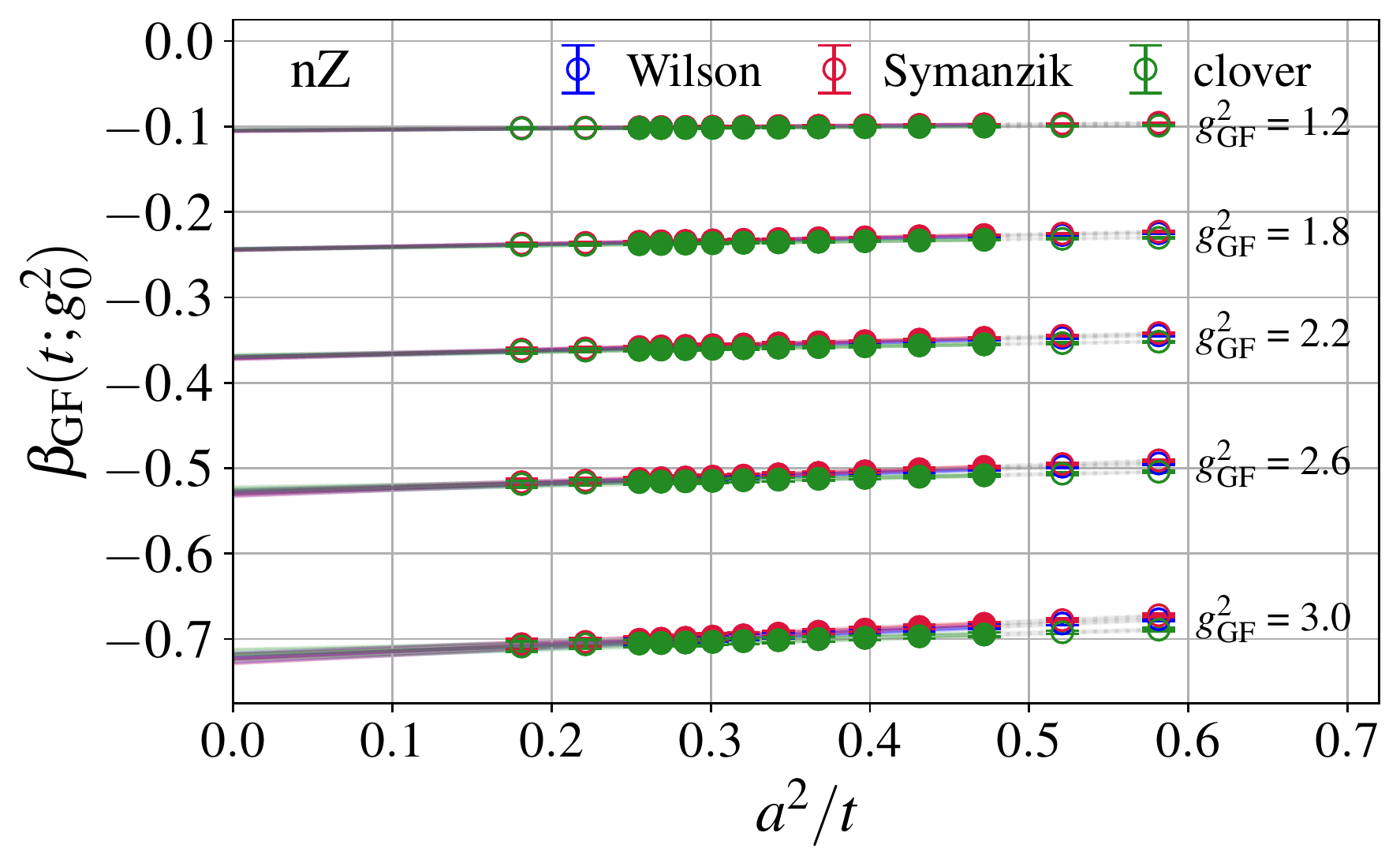}\\
    \includegraphics[width=0.95\columnwidth]{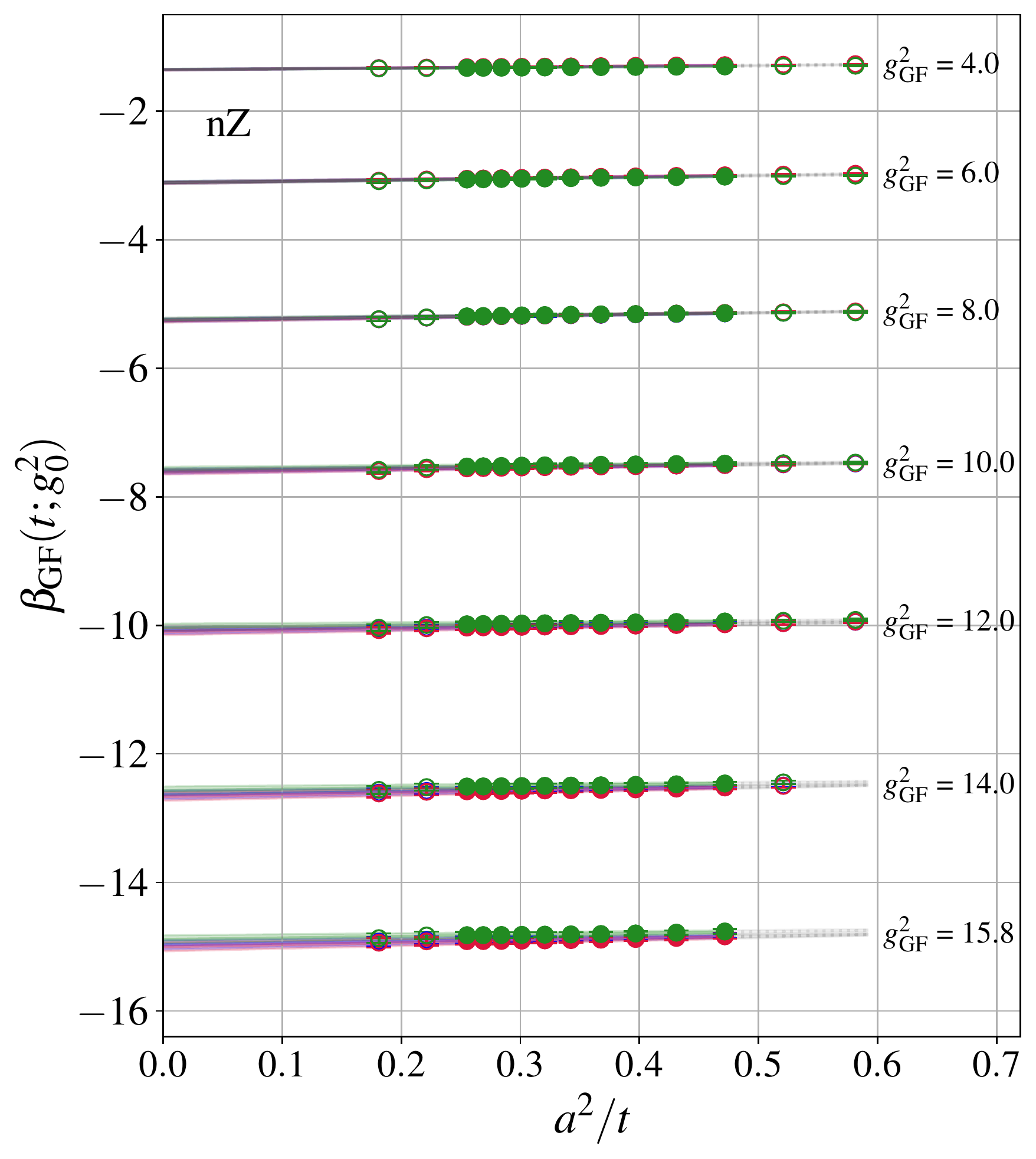}
    \caption{Continuum $a^2/t \to 0$ extrapolation between $\gGF=1.2$ to 15.8. The results are shown in two separate panels to accommodate the increasingly faster running of the coupling. In all cases, we show all three operators with Zeuthen flow after tln improvement, though the different operators overlap and are barely distinguishable in the plot. The open symbols are not included in the extrapolation fit. They are shown to illustrate the linear behavior of the data even outside the region used in the fit.}
    \label{fig:cont_extrap}
\end{figure}
\subsection{Continuum limit extrapolation}\label{subsec:cont_lim}
To obtain the continuous $\beta$ function in the continuum limit, we need to perform one last step, the $a^2/t \to 0$ continuum extrapolation. We use a linear extrapolation in $a^2/t \to 0$ and the flow time $t/a^2$ has to be chosen large enough for the RG flow to reach the renormalized trajectory where a linear extrapolation describes the remnant cutoff effects (within statistical uncertainties). We also have to limit the maximum flow time because the infinite volume extrapolation is reliable only when the finite volume corrections follow the leading order $t^2/L^4$ scaling form. In practice, we vary $t_{\mathrm{min}}/a^2$ and $t_{\mathrm{max}}/a^2$ and monitor the quality of the linear extrapolation.
In Fig.~\ref{fig:cont_extrap} we show examples of the continuum extrapolation from our weakest available  coupling of $\gGF$ up to the $t_0$ scale of $\gGF(t_0)=0.3\, \mathcal{N}\approx 15.8$. In all cases, we use $t/a^2 \in[2.0, 4.0]$ but show additional data points both at smaller and larger flow times to illustrate the linear behavior in $t/a^2$.

It is worth pointing out that the flow time is technically a continuous variable and the corresponding $\betaGF(\gGF)$ values are highly correlated.
We obtain the final continuum limit prediction by selecting a discrete set of $t/a^2$ values. For this set  we perform an uncorrelated fit and estimate its statistical uncertainty by repeating this fit using the central values shifted by $\pm1\sigma$. This avoids the complication of inverting a poorly conditioned correlation matrix and ensures we are not underestimating the statistical uncertainty. 

\subsection{The nonperturbative \texorpdfstring{$\betaGF$}{betaGF} function}\label{subsec:betafn}
\begin{figure}[tb]
    \centering
    \includegraphics[width=\columnwidth]{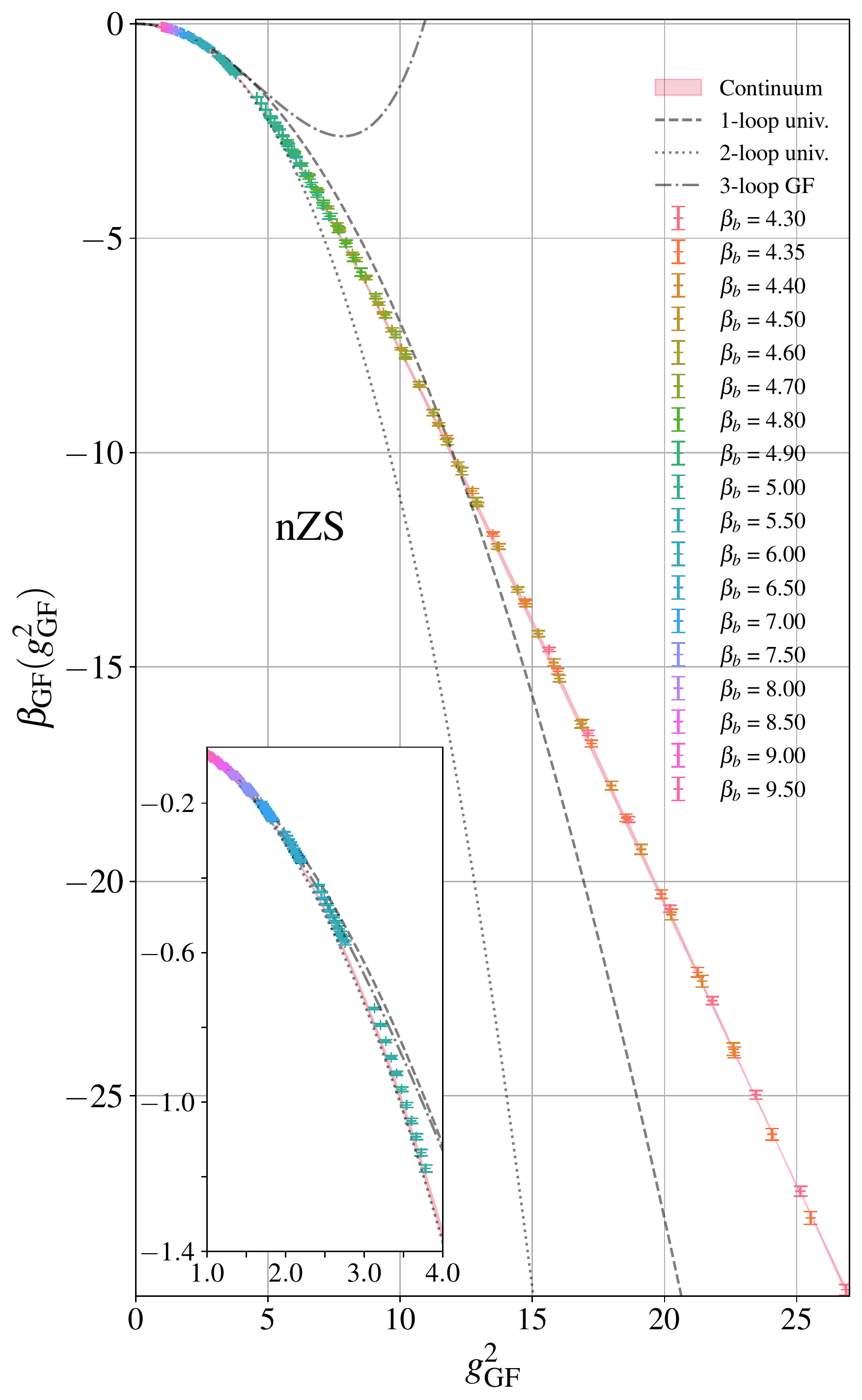}
    \caption{The predicted $\betaGF$ function (salmon colored band) overlayed with the infinite volume extrapolated data at different bare coupling $\beta_b$ (colored data points) for our main analysis based on nZS for flow times $t/a^2\in[2.0, 4.0]$,  separated by $\Delta t /a^2 = 0.2$. The insert magnifies the weak coupling region. The nZS combination shows very little cutoff dependence and the raw lattice data  sit on top of the continuum extrapolated value. }
    \label{fig:final_bt}
\end{figure}

Figure \ref{fig:final_bt} shows our result for the nonperturbative $\betaGF$ function for pure gauge SU(3) Yang-Mills in the gradient flow scheme with statistical errors only. In addition to the continuum limit prediction shown with a salmon-colored  band, the plot also shows the infinite volume extrapolated nZS lattice data at flow times $t/a^2\in[2.0, 4.0]$ where, for better visibility, we ``thin'' the data and only show every fifth data point, i.e.~flow time values are separated by $\Delta t /a^2 = 0.2$. The nZS combination shows very little cutoff dependence and the raw lattice data  sits on top of the continuum extrapolated value.  
Overall, our results  span the coupling range from $\gGF \gtrsim 1.2$  up to $\gGF\lesssim 27$, well into the confining regime of the system.

\subsection{Systematic uncertainties of the \texorpdfstring{$\betaGF$}{betaGF} function}\label{subsec:sysbetafn}

\begin{figure}[tb]
    \centering
    \includegraphics[width=\columnwidth]{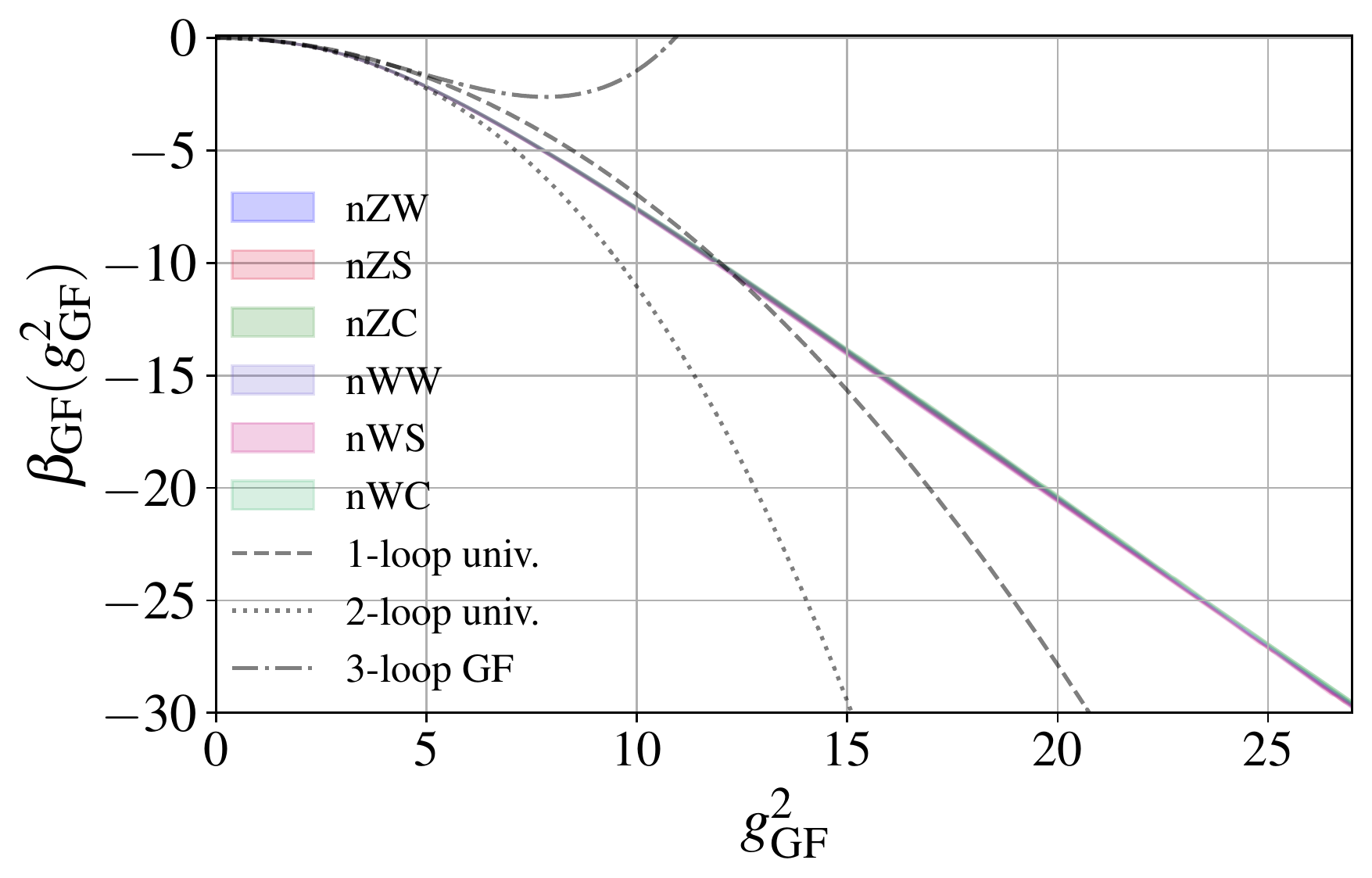}
    \caption{Comparison of different continuum limit results obtained for tln improved flow operator combinations. All results overlap and form a single band.}
   \label{fig:flow_op_comp}
\end{figure}

\begin{figure}[tb]
    \centering
    \includegraphics[width=\columnwidth]{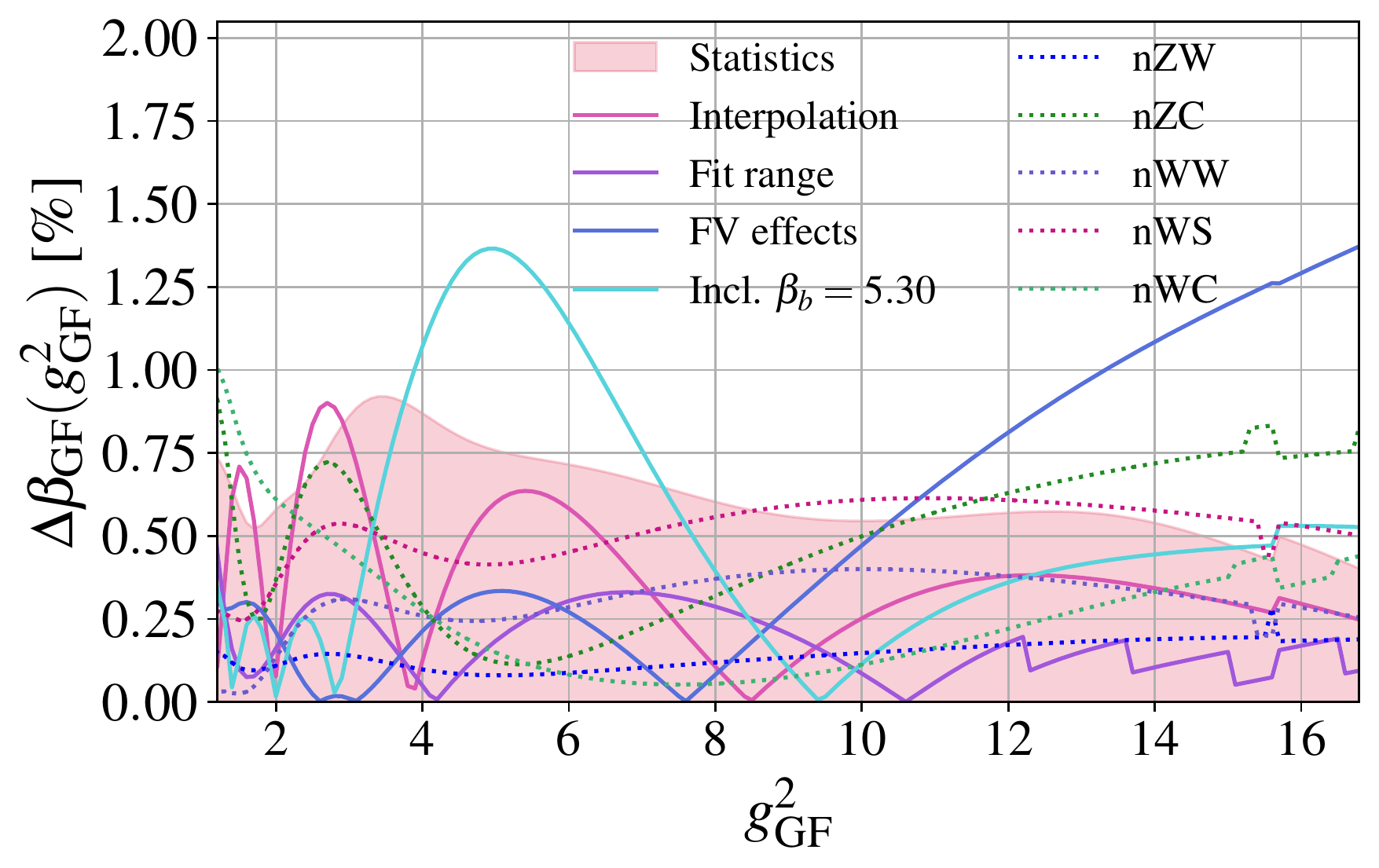}
    \caption{ Systematic uncertainties with respect to our main analysis based on nZS. By varying different parts of our analysis one after the other, we calculate the relative changes of the central value and compare the size of the different systematic uncertainties (colored lines) to our statistical uncertainty (salmon-colored band).}
   \label{fig:syst_errs}
\end{figure}

In addition to the statistical uncertainties shown for the final result of our $\betaGF$ function in Fig.~\ref{fig:final_bt}, we check for systematic effects by considering variations of our preferred nZS analysis. We discuss the different variations below and show the outcome as relative changes w.r.t.~our nZS analysis in Fig.~\ref{fig:syst_errs}. 
  \begin{itemize}
  \item For the interpolation in $\gGF$ we use a ratio of polynomials similar to Pad\'e approximation. We alter the functional form by changing the order of the polynomials in Eq.~(\ref{eq:pade}) from $N=4$ to $N=2$ or 6, but find this  is only a subleading effect on our resulting $\betaGF$ function. In Fig.~\ref{fig:syst_errs} we show the (larger) effect by taking the absolute difference between interpolations based on $N=4$ and $N=2$.
    \item Our simulations at $\beta_b=5.30$ sit right at the deconfined/confined transition and we observe poor $p$-values when performing the infinite volume extrapolation at this bare gauge coupling. We therefore discarded $\beta_b=5.30$ from our main analysis, but we use it now to check for systematic effects by including it as an additional data point in a sensitive region of the $\betaGF$ function. As expected, the impact of adding $\beta_b=5.30$ is largest around $\gGF\sim 5$. 
    \item To validate our infinite volume extrapolation, we consider two variations: 
    \begin{enumerate}
    \item We drop the smallest volume and perform a linear fit to our three largest volumes. 
    \item We note that in Fig.~\ref{fig:inf_vol_extrap} our largest volume $L/a=32$ ($L/a=48$) for $\beta_b<5.00$ ($\beta_b\ge 5.00$) is very close to the extrapolated infinite volume value. Therefore we simply repeat our analysis using only the largest volumes. 
    \end{enumerate} 
    Both variations result in comparable uncertainties, but just using the largest volumes has a slightly larger effect. Hence we use the latter to estimate finite volume effects. As can be seen in Fig.~\ref{fig:syst_errs} this could be our dominant systematic uncertainty for strong couplings $\gGF\gtrsim 11$.
    \item We test the continuum limit extrapolation by varying the range of the flow times entering the linear fit in $a^2/t$ at fixed values of $\gGF$. Keeping $t_\text{max}/a^2=4.0$ fixed, we vary $t_\text{min}/a^2$ from 1.52 to 2.0. Similarly, we vary $t_\text{max}/a^2$ from 4.0 to 5.0 while $t_\text{min}/a^2=2.0$ is kept fixed. 
    Comparing these to our preferred analysis
    we see at most a variation of $\mathcal{O}(0.3\%)$ in the central value of our $\betaGF$ function. 
    We show the maximum of these variations as ``fit range'' uncertainty in Fig.~\ref{fig:syst_errs}. Further, we repeat the fits using our preferred fit range but reduce the number of data points fitted by increasing the separation in flow time. As expected this variation results only in minuscule changes and can be neglected. 
    \item In the continuum limit different flows and operators should predict the same renormalized  $\betaGF$ function. Qualitatively this is the case, as Fig.~\ref{fig:flow_op_comp} demonstrates. There we show six different tln improved flow-operator combinations to determine $\betaGF$ which all sit on top of each other forming nearly a single band. Looking at the relative changes in Fig.~\ref{fig:syst_errs} we do, however, see deviations of $\mathcal{O}(0.8\%)$. Since for couplings in the range $7\lesssim \gGF\lesssim 11$ that effect is larger than other systematic effects, we conservatively include these variations when obtaining our systematic uncertainty. 
    \item We further check for consistency by analyzing our data without using tln improvement. We find that removing the tln improvement increases the discretization errors noticeably, but within the larger uncertainties, the results are consistent with our preferred analysis.
  \end{itemize}

\begin{figure}[tb]
    \centering
    \includegraphics[width=\columnwidth]{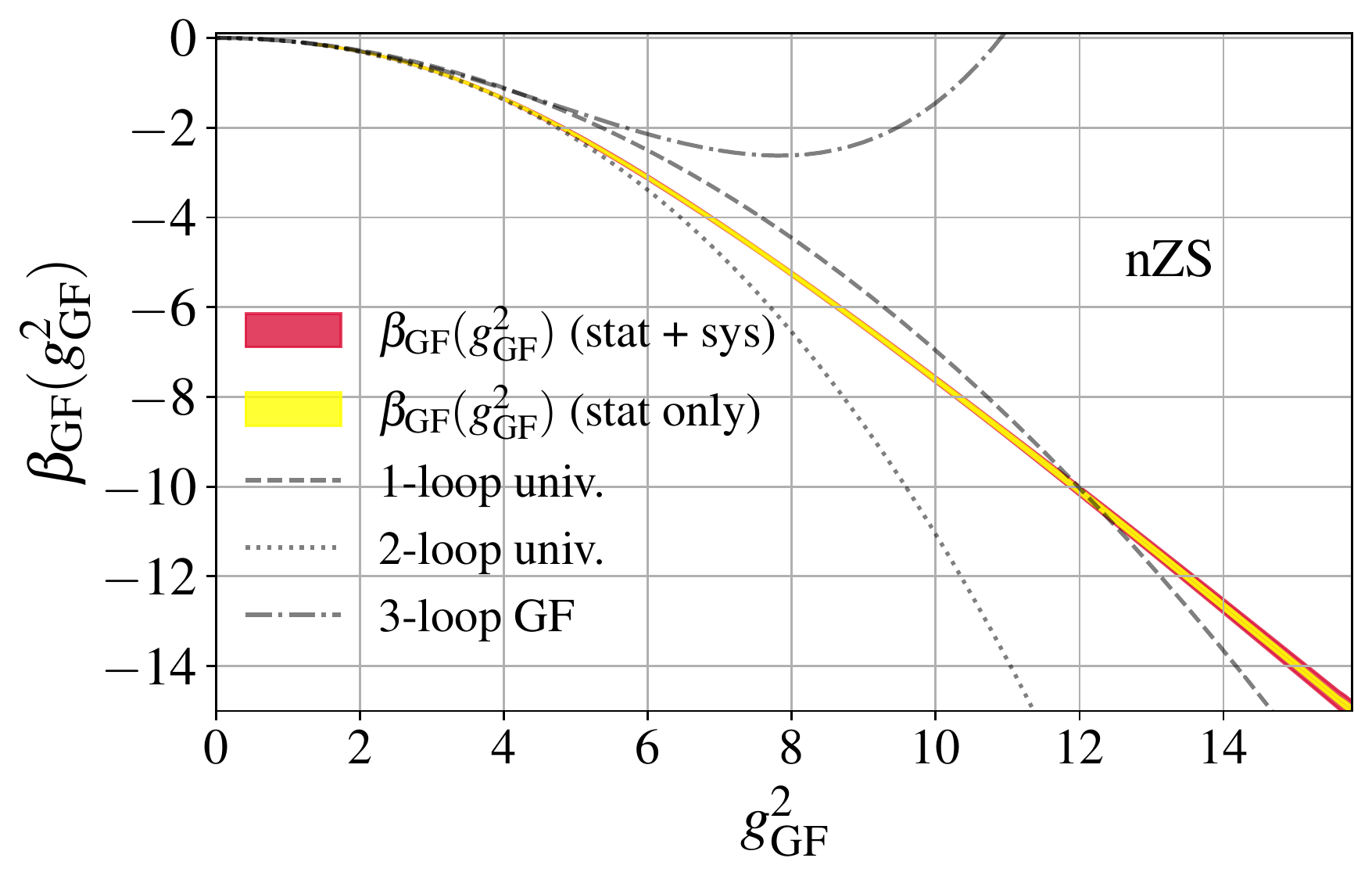}
    \caption{Final result for $\betaGF$ as a function of $\gGF$ for the coupling range relevant to determine the $\Lambda$ parameter. The yellow inner band shows only the statistical uncertainty, whereas the red outer band shows the combined statistical and systematic uncertainties. }
   \label{fig:finalBeta}
\end{figure}

Using the information compiled in Fig.~\ref{fig:syst_errs}, we obtain the total uncertainty of our $\betaGF$ function by adding our statistical error and the largest systematic effect at each $\gGF$ value in quadrature. Our final result for the $\betaGF$ function for the range of $\gGF$ relevant to determine $\sqrt{8t_0}\LambdaGF$ (see below) is shown in Fig.~\ref{fig:finalBeta}.\footnote{Our final result for $\betaGF$ is also provided as an ASCII file \cite{supp}.}

\subsection{\texorpdfstring{$\betaGF$}{betaGF} in the strong coupling regime}\label{subsec:strong_coup_reg}

As can be seen in particular in Fig.~\ref{fig:flow_op_comp}, the $\betaGF$ function in the strong coupling regime ($\gGF \gtrsim 13$) is  approximately linear. Studying the derivative $\mathrm{d}\betaGF/\mathrm{d}\gGF$, we observe a plateau within errors in the range $20 \lesssim \gGF \lesssim 27$. In that  range  we can approximate $\betaGF(\gGF) \approx c_0 + c_1 \gGF$ with $c_0 = 5.8(3)$, $c_1=-1.32(1)$. This is very different from the perturbative prediction that would suggest a polynomial form with terms of $g_{\mathrm{GF}}^4$ and higher order. Clearly, nonperturbative effects are at play here. 
We  may use the above observation to predict the flow time dependence of the flowed energy density $\vev{t^2 E(t)} \sim a_0 + a_1  \big(t / \tilde{t} )^{-c_1}$, where $a_0=-c_0/c_1\approx 4.4(2)$ and $\tilde{t}$ is an integration constant.
We do not have a physical explanation for this behavior, though it is tempting to think that  the form of $\vev{E(t)}$ as $t\rightarrow\infty$  is related to instantons, the only objects in the vacuum after UV fluctuations are removed by the flow. We mention that Refs.~\cite{Nakamura:2021meh,Schierholz:2022wuc,Ryttov:2007cx, Chaichian:2018cyv} hypothesize that in confining systems the RG $\beta$ function is linear in the confining, strongly coupled regime. 
 References \cite{Nakamura:2021meh,Schierholz:2022wuc} attempt to solve the strong CP problem by connecting $\vev{E(t)}$ to the instanton density at large flow time. However,  this work considers only one bare gauge coupling, does not take the continuum limit and uses 
 much larger flow times than we do on similar volumes. While our results are consistent with a linear $\beta$ function  in the strong coupling range, our  slope $c_1=-1.32(1)$ is significantly smaller than -1 obtained in Refs.~\cite{Nakamura:2021meh,Schierholz:2022wuc}. Moreover, we observe a nonzero intercept. 
 It would be interesting to understand the nonperturbative origin of the linearity of the $\betaGF$ function in the strong coupling confining region both in pure Yang-Mills systems and, if it persists, with dynamical fermions.

\section{The \texorpdfstring{$\Lambda$}{Lambda} parameter}\label{sec:lmbda}

\subsection{Matching the perturbative regime}\label{subsec:eff_bf}

\begin{figure}[tb]
    \centering
    \includegraphics[width=\columnwidth]{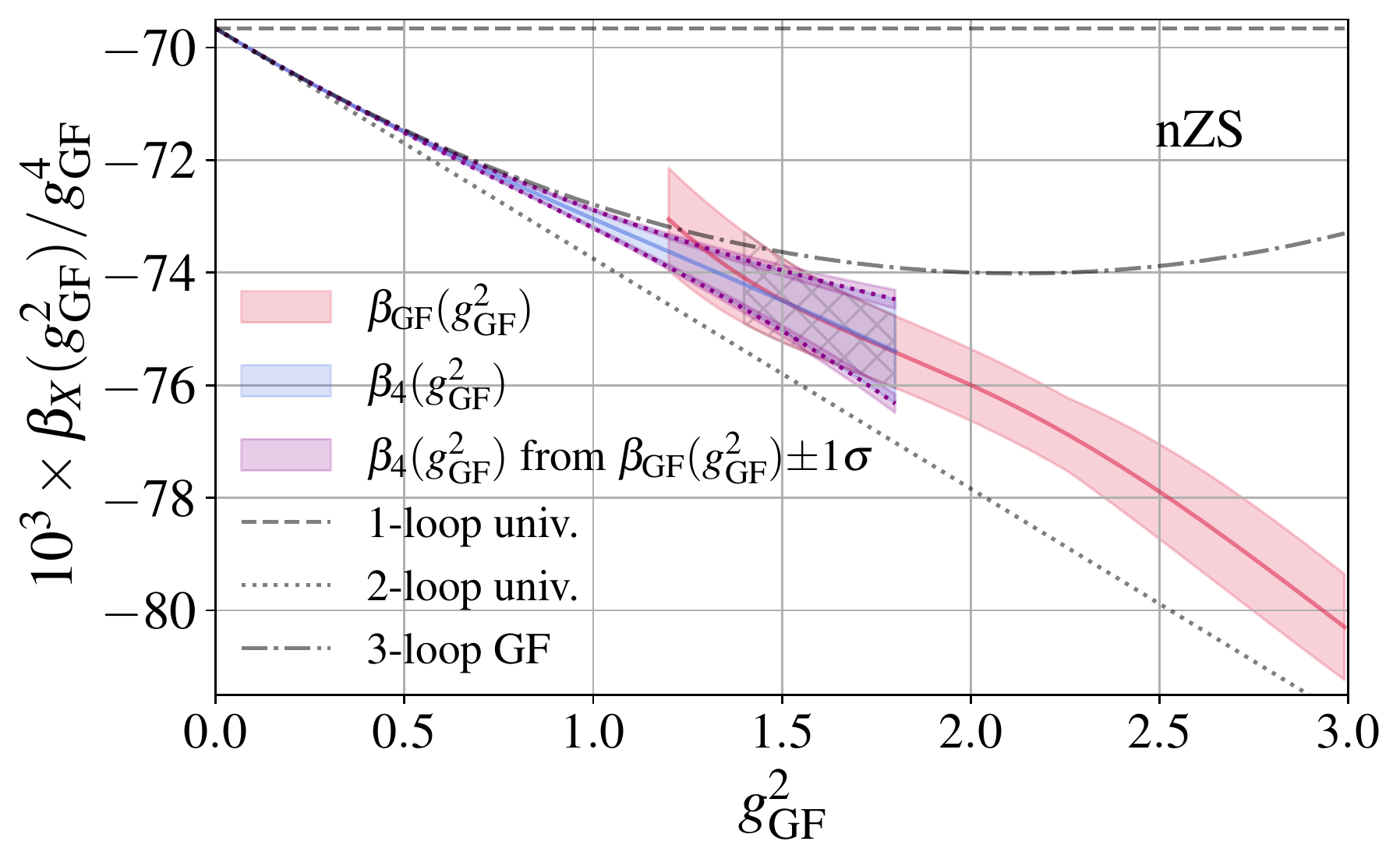}
    \caption{$\betaGF(\gGF)/g^4_{\mathrm{GF}}$ in the weak coupling region. The  salmon-colored band shows  our nonperturbatively determined $\betaGF$ with combined statistical and systematic uncertainties. We match to the 3-loop GF function using Eq.~(\ref{eqn:bt_fn_prmtrzn}) in the range $\gGF \in [1.4,1.8]$ indicated by the grey hatched area. Shifting our nonperturbative values by $\pm1\sigma$, we obtain the magenta bands providing the upper and lower limits of the resulting matched function shown in blue.} 
    \label{fig:effbeta}
\end{figure}

\begin{figure}[tb]
    \centering
    \includegraphics[width=1.0\columnwidth]{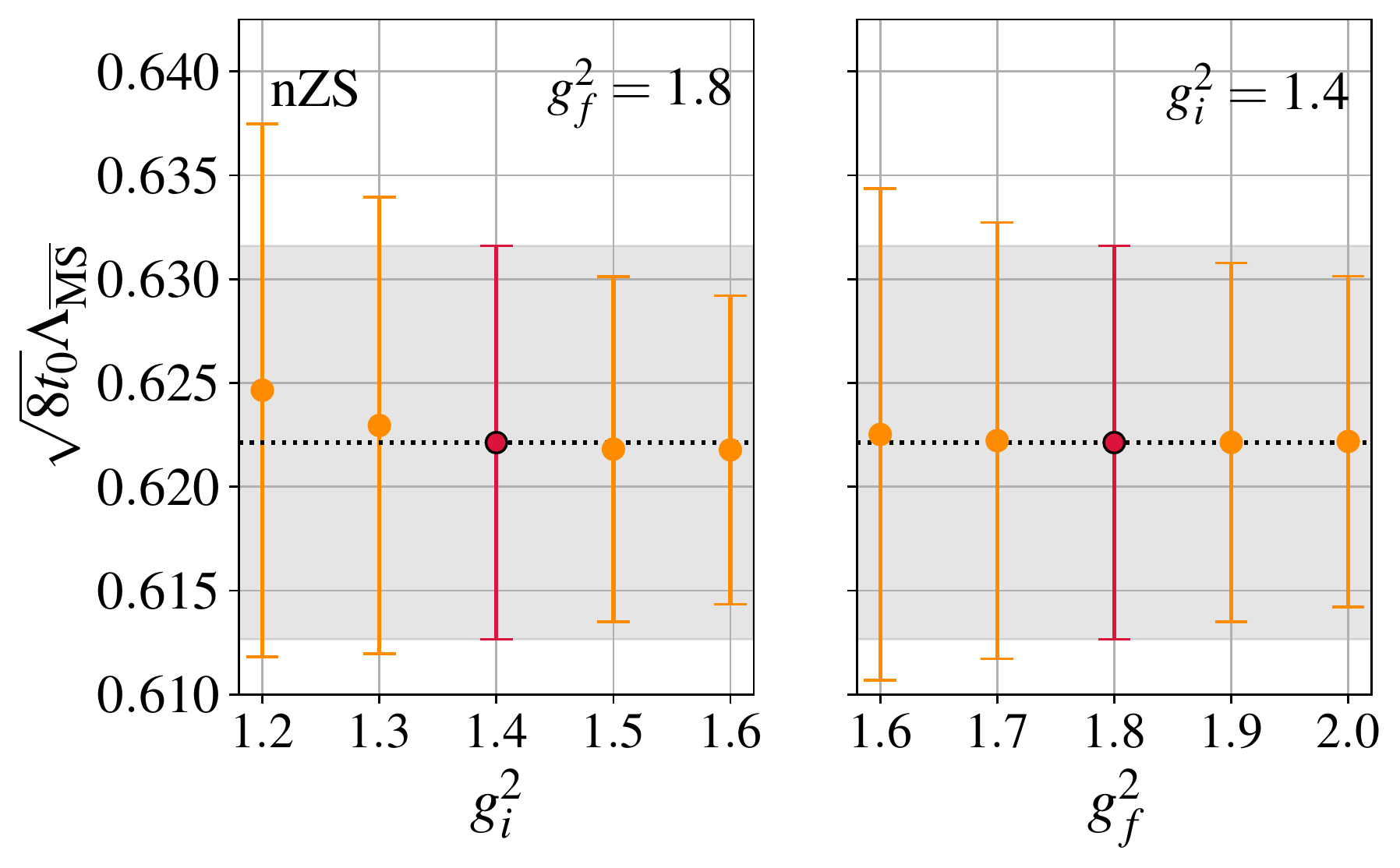}
    \caption{Systematic uncertainty in our perturbative matching procedure due to choosing $g_i^2$ or $g_f^2$.  The value for our preferred choices for $g_i^2$ and $g_f^2$ is highlighted in red.}
   \label{fig:result_for_lambda_param}
\end{figure}

In order to determine the $\Lambda$ parameter according to Eq.~(\ref{eqn:lambda_parameter}), we need to know $\betaGF(\gGF)$ down to $\gGF = 0$.
In the weak coupling regime we expect to recover the perturbatively predicted  $\beta$ function. In Fig.~\ref{fig:effbeta} we show $\betaGF/g^4_{\textrm{GF}}$ to emphasize the weak coupling behavior of our nonperturbative result. The numerical data, shown by the salmon-colored band, is close to the 3-loop GF perturbative curve at our weakest gauge coupling $\gGF\approx$ 1.2, but it does not yet connect smoothly.
To remedy this  limitation we extend our numerically determined $\betaGF$ function to the region below our weakest coupling data point $(\gGF \approx$ 1.2) using the parameterization
\begin{equation}\label{eqn:bt_fn_prmtrzn}
    \beta_{4}(\gGF) \equiv -g^4_{\mathrm{GF}}\Big(b_0 + b_1 g^2_{\mathrm{GF}} + b_2 g^4_{\mathrm{GF}} + b_p g^6_{\mathrm{GF}}\Big),
\end{equation}
where $b_0$, $b_1$, $b_2$ are the 1-, 2- and 3-loop GF perturbative coefficients from Ref.~\cite{Harlander:2016vzb} and $b_p$ is free. We  determine $b_p$  by   integrating the inverse $\beta$ function
\begin{equation}\label{eqn:flow_time_rat}
    \int_{g^2_i}^{g^2_f} \!\!\!\!\! \mathrm{d}x \; \beta^{-1}(x),
\end{equation}
using both our numerically determined $\betaGF$ function and also using $\beta_{4}$ parametrized by $b_p$ for $\beta(x)$.  We determine $b_p$ by equating these two values. The integration limits are set to cover the regime where we attempt to match $\betaGF$ and $\beta_{4}$. In the example shown in  Fig.~\ref{fig:effbeta}, we choose to match in the range $g^2_i=1.4$, $g^2_f=1.8$. In Fig.~\ref{fig:result_for_lambda_param} we demonstrate that varying   $g_i^2$ and $g_f^2$ has negligible impact on our value of $\sqrt{8t_0}\Lambda_{\overline{\text{MS}}}$. To account for the combined statistical and systematic uncertainty of our nonperturbative $\betaGF$ function, we repeat the matching shifting the central values by $\pm1\sigma$. That way we obtain the purple bands resulting in the upper  and lower end of the blue band connecting our nonperturbative result to $\gGF=0$. 

\subsection{Calculating the \texorpdfstring{$\Lambda$}{Lambda} parameter}\label{subsec:lmbda_calc}
 
The final step of this analysis is to calculate $\sqrt{8 t_0} \LambdaGF$ by integrating Eq.~(\ref{eqn:lambda_parameter}) using the combination of  our nonperturbative $\betaGF(\gGF)$ for $\gGF\ge 1.4$ and the matched $\beta_4(\gGF)$ function for $\gGF<1.4$.  The upper integration limit is set according to $\gGF(t_0)\approx 15.8$. Our prediction is
  \begin{align}
    \sqrt{8 t_0} \Lambda_{\mathrm{GF}} = 1.164(19),
  \end{align}
  where the error accounts for the uncertainty of our $\betaGF$ function as well as the uncertainty encountered due to the matching procedure.  As a last step, we convert our prediction to the $\overline{\text{MS}}$ scheme using the exact 1-loop relation and find
  \begin{align}
    \sqrt{8 t_0} \Lambda_{\overline{\text{MS}}} = 0.622(10).
  \end{align}

\subsection{Comparison of  \texorpdfstring{$\sqrt{8t_0}\Lambda_{\MSbar}$}{sqrt(8t0) LambdaMS} determinations}\label{subsec:lmbda_comp}
\begin{figure}[tb]
    \centering
    \includegraphics[width=1.0\columnwidth]{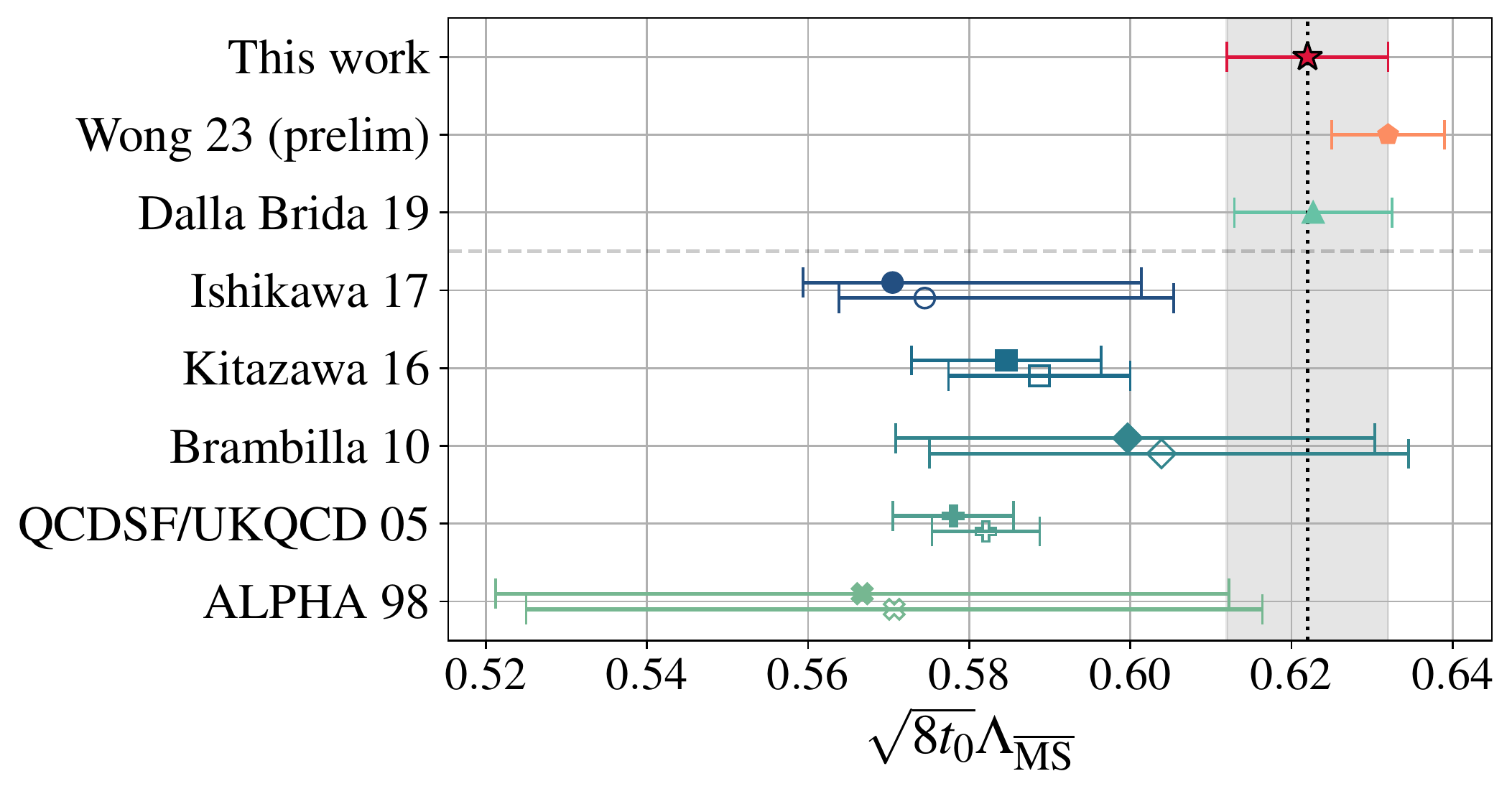}
    \caption{Comparison of our result for $\sqrt{8 t_0}\LambdaGF$ ($\bigstar$) to the preliminary result by Wong et al.~\cite{Wong:2023jvr} ($\pentagonblack$) and Dalla Brida/Ramos \cite{DallaBrida:2019wur} ($\blacktriangle$). In addition we show  values  for $r_0\Lambda_{\MSbar}$ which enter the FLAG 2021 averages: ALPHA 98 \cite{Capitani:1998mq} ($\times$), QCDSF/UKQCD 05 \cite{Gockeler:2005rv} ($+$), Brambilla~10 \cite{Brambilla:2010pp} ($\blacklozenge$), Kitazawa 16 \cite{Kitazawa:2016dsl} ($\mdblksquare$), and Ishikawa 17 \cite{Ishikawa:2017xam} ($\mdblkcircle$). These values are converted to $\sqrt{8 t_0}\Lambda_{\mathrm{GF}}$ using $\sqrt{8t_0}/r_0$ from \cite{Luscher:2010iy} (open symbols) or Ref.~\cite{DallaBrida:2019wur} (filled symbols).}
    \label{fig:result_for_lambda_param_cmp}
\end{figure}

The Yang-Mills $\Lambda$ parameter has been studied previously using different approaches. Only the recent gradient flow studies in Refs.~\cite{DallaBrida:2019wur,Wong:2023jvr} directly predict the combination $\sqrt{8 t_0}\Lambda_{\mathrm{\MSbar}}$. In addition, we compare our value to determinations of $r_0\Lambda_{\MSbar}$ listed by the flavor lattice 
averaging group (FLAG) \cite{FlavourLatticeAveragingGroupFLAG:2021npn} to meet the quality criteria to enter the average. These determinations are obtained using Schr\"odinger functional step-scaling methods \cite{Capitani:1998mq,Ishikawa:2017xam}, Wilson loops \cite{Gockeler:2005rv,Kitazawa:2016dsl}, or the short distance potential \cite{Brambilla:2010pp}. 
We use the values quoted by FLAG 2021 for $r_0\Lambda_{\MSbar}$  and  
convert them to $\sqrt{8t_0}\Lambda_{\MSbar}$ using $\sqrt{8t_0}/r_0 = 0.948(7)$ \cite{Luscher:2010iy} (open symbols) or  $\sqrt{8t_0}/r_0 = 0.9414(90)$ \cite{DallaBrida:2019wur} (filled symbols). 
Following the FLAG convention, we refer to the different results in Fig.~\ref{fig:result_for_lambda_param_cmp} using either the name of the first author or, if applicable, the name of the collaboration and the two-digit year.

Given the spread in the values of $\sqrt{8t_0}\Lambda_{\MSbar}$, further scrutiny and understanding are needed before obtaining an average. 
We note, however, that the three most recent predictions are all mutually consistent. The high-precision result of Ref.~\cite{DallaBrida:2019wur} was re-affirmed in Ref.~\cite{Nada:2020jay} using an alternative approach with better control over the continuum extrapolation.  The estimate given in Ref.~\cite{Schierholz:2022wuc} is also consistent with these predictions.
A possible source of difference to the older determinations is the conversion of $r_0$ to $\sqrt{8t_0}$.

\subsection{Nonperturbative matching of different schemes}\label{subsec:npmatching}

In our analysis, a considerable systematic uncertainty arises from the weak coupling limit $\gGF \lesssim 1.4$. Our gradient flow setup is not efficient at weak coupling. It would be more economical to use data from existing calculations, e.g.~the high precision Schr\"odinger functional data of Ref.~\cite{DallaBrida:2019wur} in the $0<\gGF<1.4$ regime and match it nonperturbatively to our data.

Such a matching requires finding the relation between our $\gGF$ coupling and the coupling $g^2_\mathcal{S}$ of another scheme $\mathcal{S}$ expressed as $\gGF=\phi\big(g^2_\mathcal{S}\big)$.  The relation of the corresponding $\beta$ functions can be obtained using the chain rule applied to the derivative of $\gGF$ with respect to $\mu^2$, which leads to the simple relation
\begin{equation}
\frac{\betaGF\left( \phi(g_\mathcal{S}^2) \right)}{\phi^\prime\big(g_\mathcal{S}^2\big)} = \beta_{S}\big(g_\mathcal{S}^2\big),
\label{eq:npmatch}
\end{equation}
where $\phi'\big(g^2_\mathcal{S}\big) \equiv \mathrm{d} \phi\big(g^2_\mathcal{S}\big)/\mathrm{d}g^2_\mathcal{S}$.
Parametrizing $\phi$ as a polynomial
\begin{equation}\label{eqn:func_dep}
    \phi(x) \approx x + x^2\sum_{n=0}^{N_p-1} c_{i}x^{i},
\end{equation}
turns  Eq.~(\ref{eq:npmatch}) into a straightforward fitting problem with $N_p$ undetermined coefficients. The only constraint is to identify and use the renormalized coupling range in the fit where the two schemes overlap. Such a nonperturbative matching and combination of different schemes could lead to a significantly improved prediction. Although we do not explore this method in the present analysis, it is worth considering in the future.

\section{Discussion}\label{sec:discussion}

In this paper, we present a nonperturbative determination of the renormalization group $\beta$ function for the pure gauge Yang-Mills action. Using the gradient flow based continuous RG $\beta$ function, we present results for a wide range of values of the renormalized running coupling. Our results span the range of the perturbative weak coupling region $\gGF\approx 1.2$ up to the strongly coupled regime at $\gGF \approx 27$.  This showcases the advantage of the continuous RG $\beta$ function because the continuous infinite volume $\beta$ function can be extended without limitation to the confining region. We also demonstrate the effectiveness of tree-level  improvement of the gradient flow even in the strong coupling regime.

We investigate various sources of systematical uncertainties.
For most of the $\gGF$ range covered, our statistical uncertainties are around 0.6\%. In the weak coupling region, statistical and systematic errors are of similar size. At intermediate to strong coupling, we observe an increase in the systematic errors to 1.5\% due to enhanced finite volume effects.

While in the weak coupling our results are close to the perturbative values, we observe in the confining regime that the GF $\beta$ function depends  approximately linearly on the running coupling, implying a scaling relation of the flowed energy density $\vev{t^2 E(t)} \sim a_0 + a_1 \big(t/\tilde{t}\big)^{-c_1}$ with exponent $c_1\approx -1.32(1)$. This observation could be related to the topological structure of the vacuum, a possibility that  warrants further investigation.

In the weak coupling regime, we are able to match our numerical results to the 3-loop GF $\beta$ function by extending the perturbative expression with a single $g^{10}_{\mathrm{GF}}$ term. This matching allows us to predict the $\Lambda$ parameter in the GF scheme. Using the perturbatively determined relation of the GF coupling $\gGF$ and the $\overline{\textrm{MS}}$ coupling, we obtain $\sqrt{8 t_0} \Lambda_{\MSbar}=0.622(10)$, where the error  combines statistical and systematic  uncertainties. This value is in good agreement with recent direct determinations of $\sqrt{8 t_0} \Lambda_{\MSbar}$ \cite{Wong:2023jvr,DallaBrida:2019wur}. 

A significant source of the systematic uncertainties in determining $\sqrt{8 t_0} \Lambda_{\MSbar}$ arises from the weak coupling regime. We outline a nonperturbative matching procedure to combine existing high-precision data in the weak coupling and our nonperturbative $\betaGF$ function that extends into the confining regime even beyond $\gGF(t_0)$. Such a combined determination could lead to a sizable reduction of the uncertainties on $\sqrt{8 t_0} \Lambda_{\MSbar}$.


\begin{acknowledgments}
We are very grateful to Peter Boyle, Guido Cossu, Anontin Portelli, and Azusa Yamaguchi, who
develop the \texttt{GRID} software library providing the basis of this work and who assisted us in installing
and running \texttt{GRID} on different architectures and computing centers. We benefited from many comments and discussions during the ``Gradient Flow in QCD and Other Strong Coupled Gauge Theories'' Workshop at the European Center for Theoretical Studies in Nuclear Physics and Other Related Areas (ECT*), Trento, Italy, March 20-24, 2023. A.H. acknowledges support by DOE Grant No.~DE-SC0010005. This material is based upon work supported by the National Science Foundation Graduate Research Fellowship Program under Grant No.~DGE 2040434.

Computations for this work were carried out in part on facilities of the USQCD Collaboration, which are funded by the Office of Science of the U.S.~Department of Energy, the RMACC Summit supercomputer \cite{UCsummit}, which is supported by the National Science Foundation (awards No.~ACI-1532235 and No.~ACI-1532236), the University of Colorado Boulder, and Colorado State University, and the OMNI cluster of the University of Siegen.  We thank  BNL, Fermilab, the University of Colorado Boulder, the University of Siegen, the NSF, and the U.S.~DOE for providing the facilities essential for the completion of this work.
\end{acknowledgments}

\bibliography{./BSM}
\end{document}